\documentclass[letterpaper,twocolumn,10pt]{article}
\usepackage{usenix,epsfig,endnotes}

\usepackage{microtype}
\usepackage{url}
\usepackage{tikz}
\usepackage{subfig}
\usepackage{multirow}
\usepackage{booktabs}
\usepackage{graphicx}
\usepackage{xspace}
\usepackage{algorithm}
\usepackage[noend]{algpseudocode}
\usepackage{balance}
\usepackage{amsmath}
\usepackage{mathtools}
\usepackage{amsfonts}
\usepackage{amsthm}
\usepackage{threeparttable}
\usepackage{xcolor}
\usepackage{enumitem,kantlipsum}
\usepackage{siunitx}
\usepackage[compact]{titlesec}
\usepackage[nosort]{cleveref} 
\usepackage{url}

\setlist{nosep}
\newtheorem{definition}{Definition}

\newcommand{\Sys}{Parcae\xspace}
\newcommand{\SysSched}{ParcaeScheduler\xspace}
\newcommand{\SysAgent}{ParcaeAgent\xspace}
\newcommand{\SysAgents}{ParcaeAgents\xspace}
\newcommand{\SysPS}{ParcaePS\xspace}
\newcommand{\LPT}{\textproc{Liveput}\xspace}

\newcommand{\liveput}{liveput\xspace}
\newcommand{\throughput}{throughput\xspace}

\newcommand{\TPT}{\textproc{Throughput}\xspace}

\newcommand{\shepherdcommentout}[1]{}
\newcommand{\captionvspace}{-1em}
\newcommand{\commentout}[1]{}

\renewcommand{\paragraph}[1]{
     \textit{\textbf{#1}}
 }

\DeclareMathOperator*{\argmax}{arg\,max}
\algnewcommand{\LeftComment}[1]{\Statex \(\triangleright\) #1}

\newlength{\myeqskip}  
\AtBeginDocument{%
    \setlength\abovedisplayskip{\myeqskip}%
    \setlength\belowdisplayskip{\myeqskip}%
    \setlength\abovedisplayshortskip{\myeqskip-\baselineskip}%
    \setlength\belowdisplayshortskip{\myeqskip}}

\makeatletter
\newcommand\blfootnote[1]{%
  \begingroup
  \renewcommand\thefootnote{}\footnote{#1}%
  \addtocounter{footnote}{-1}%
  \endgroup
}
\makeatother

\begin{document}

\title{\Sys: Proactive, Liveput-Optimized DNN Training on Preemptible Instances}

\author{\rm{Jiangfei Duan}$^{\ddag\spadesuit}$ \hspace{1.2em} \rm{Ziang Song}$^{\S\spadesuit}$ \hspace{1.2em} Xupeng Miao$^{\dag\spadesuit}$ \hspace{1.2em} Xiaoli Xi$^\dag$\\
 \rm{Dahua Lin}$^\ddag$ \hspace{1.2em} Harry Xu$^{\sharp}$ \hspace{1.2em} Minjia Zhang$^{\diamond}$ \hspace{1.2em} Zhihao Jia$^{\dag}$ \\
\\
Carnegie Mellon University$^\dag$ \hspace{1.2em} The Chinese University of Hong Kong$^\ddag$ \\
ByteDance$^\S$ \hspace{1.2em} University of California, Los Angeles$^{\sharp}$ \hspace{1.2em} Microsoft$^{\diamond}$
}

\date{}



\maketitle

\pagestyle{empty}

\subsection*{Abstract}
Deep neural networks (DNNs) are becoming progressively large and costly to train. This paper aims to reduce DNN training costs by leveraging preemptible instances on modern clouds, which can be allocated at a much lower price when idle but may be preempted by the cloud provider at any time.
Prior work that supports DNN training on preemptive instances employs a {\em reactive} approach to handling instance preemptions and allocations after their occurrence, which only achieves limited performance and scalability.
\blfootnote{$^\spadesuit$ Contributed equally. Work done during internships at CMU.}

We present \Sys, a system that enables cheap, fast, and scalable DNN training on preemptible instances by {\em proactively} adjusting the parallelization strategy of a DNN training job to adapt to predicted resource changes before instance preemptions and allocations really happen, which significantly reduces the cost of handling these events. 
\Sys optimizes {\em liveput}, a novel metric that measures the {\em expected} training throughput of a DNN job under various possible preemption scenarios. 
Compared to existing reactive, throughput-optimized systems, \Sys's proactive, live-optimized solution considers both the throughput of a job and its robustness under preemptions.
To optimize liveput, \Sys supports lightweight instance migration and uses an availability predictor to forecast future preemptions. It then uses a liveput optimizer to discover an optimal strategy to parallelize DNN training under predicted preemptions.
We evaluate \Sys on a variety of DNNs and preemption traces and show that \Sys outperforms existing spot-instance DNN training systems by up to 10$\times$. More importantly, \Sys achieves near-optimal performance for training large DNNs under frequent preemptions, in which case existing approaches cannot make any progress.

\section{Introduction}
\label{sec:intro}

Deep neural networks (DNNs) have surpassed human predictive performance on a spectrum of tasks, including computer vision~\cite{resnet}, natural language progressing~\cite{bert}, game playing~\cite{SilverHuangEtAl16nature}, and content generation~\cite{song2020score}.
The success of DNNs is associated with progressively increasing energy and financial costs.
For example, a single training run of GPT-3~\cite{gpt3}, a language model with 175 billion parameters, requires more than 1.5 million GPU hours and costs \$4.6 million to train on AWS even with the lowest priced GPUs~\cite{patterson2021carbon}.
While pre-trained models are publicly available and can be fine-tuned for different downstream tasks, training new models is often required for emerging applications and datasets.

Modern cloud platforms provide a variety of cheap {\em preemptible instances}, which can be leveraged to minimize the monetary cost of DNN training.
First, spot GPU instances allow users to take advantage of unused GPU capacity at a price up to 90\% lower than on-demand counterparts~\cite{spot_instance}.
Second, modern data centers generally reserve additional GPU capacity for urgent jobs, which can be allocated by other jobs in a preemptible manner~\cite{newell2021ras}.
Third, some ML systems~\cite{xiao2020antman} support opportunistically running training jobs on inference-dedicated GPUs to maximize resource utilization and preempt these training jobs when inference requests arrive.
While this paper focuses on spot GPUs, our techniques can easily generalize to other preemptible resources.

\begin{figure*}[t]
    \centering
    \includegraphics[width=1.0\linewidth]{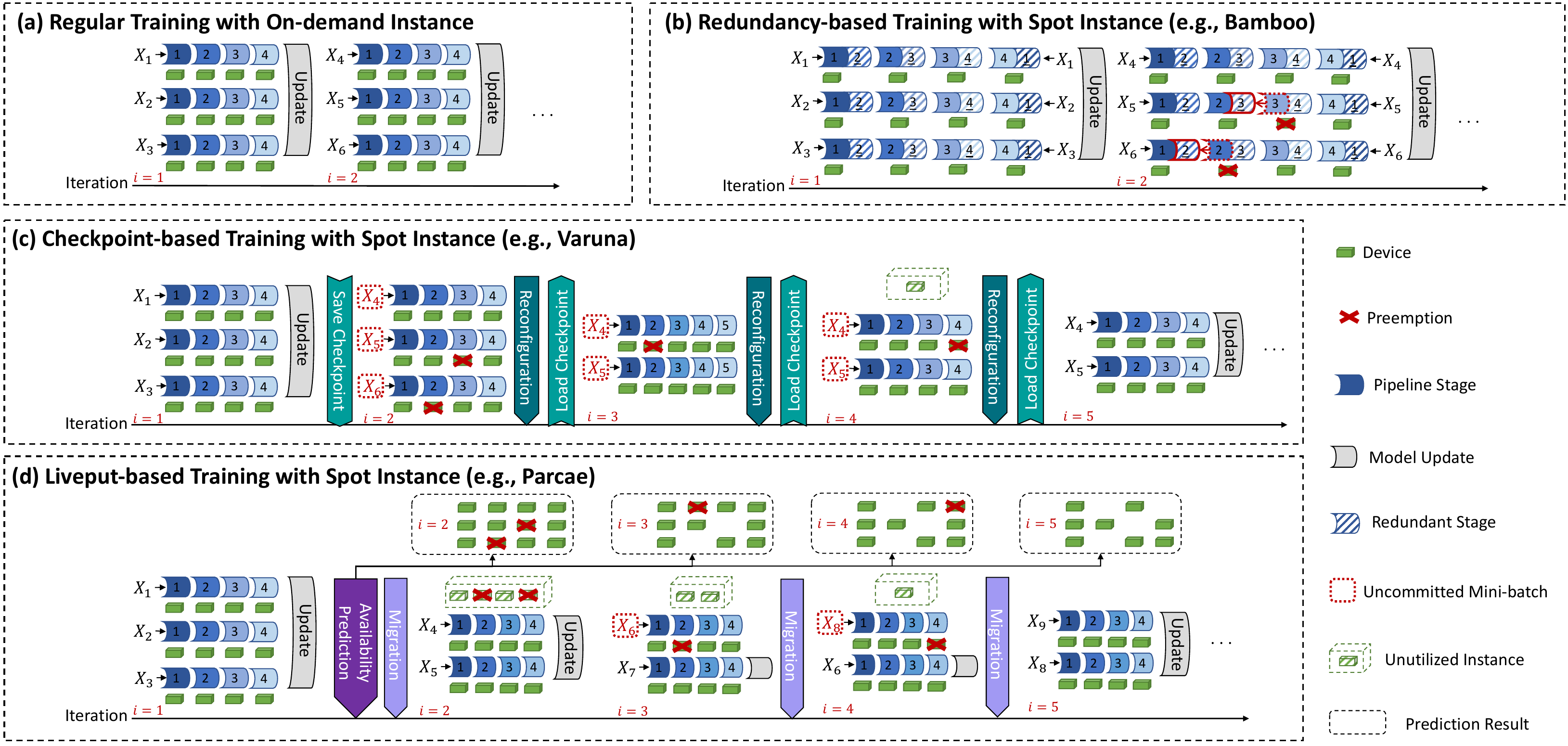}
    \vspace{\captionvspace}
    \caption{Illustration of pipelined data parallelism training over on-demand and spot instance respectively. Preempted spot instances are marked with red markers. $X_j$ represents the $j$-th mini-batch of input data.}
    \label{fig:comparison}
\end{figure*}

Existing systems that support DNN training on spot instances use a {\em reactive} approach to handling instance preemption and allocation, and can be categorized into two classes: {\em checkpoint}- and {\em redundancy-based} systems. We introduce the two categories and identify the limitations of these reactive approaches in {\em performance} and {\em scalability} when applied to DNN training on preemptible instances.

The first line of work uses {\em checkpoints} to maintain model states during training.
For example, Varuna~\cite{athlur2022varuna} periodically saves model states to persistent storage and loads the latest checkpoint back after a preemption, as shown in \Cref{fig:comparison}c.
Although Varuna offers promising training throughput when spot instances have low preemption rates, it struggles to make progress when preemptions are frequent. 
This is due to two reasons: (1) saving and loading checkpoints incur significant IO overhead, particularly as model size increases, making frequent checkpointing costly, and (2) high preemption rates cause training to frequently roll back to the last saved checkpoint, resulting in wasted computation as model updates made since the last checkpoint are lost.

The second line of work uses redundant computation to provide resilience in the presence of preemptions.
For example, as shown in \Cref{fig:comparison}b, Bamboo~\cite{bamboo} replicates DNN computations across spot instances by letting each instance in a pipeline perform normal computations over assigned DNN layers (dark boxes) and redundant computations over its successor's layers (striped boxes).
Upon an instance preemption, its predecessor has all the information (e.g., layers and activations) to continue DNN training.
Although Bamboo achieves higher training throughput than pure checkpointing-based methods when preemption rates are high, its computation efficiency can still be limited. 
This is because it is difficult to completely hide the overhead of redundant computation through pipeline bubbles, especially for large-scale models (\S\ref{eval:end_to_end}).
Additionally, storing redundant model states increases per-GPU memory consumption.
Existing redundancy-based methods such as Bamboo address this challenge by increasing pipeline depth, but this can lead to reduced computation efficiency and increased vulnerability to preemptions.

To address the performance and scalability limitations of existing approaches, this paper presents \Sys, a {\em proactive}, {\em liveput-optimized} system for DNN training on spot instances.
\Sys combines data and pipeline parallelism for DNN training on spot instances, and maintains identical semantics as on-demand training.
A key insight behind \Sys is that different strategies to parallelize DNN training exhibit {\em diverse robustness} under preemptions.
For example, a strategy with long pipelines achieves higher throughput but is more vulnerable to preemptions than a strategy with shorter pipelines.

\Sys is designed to maximize {\em preemption-aware} throughput in a proactive way. We purpose a formulation of {\em liveput} for DNN training on preemptible instances, which is the {\em expected} training throughput of a DNN job under different preemption scenarios. 
A key advantage of liveput is that it considers both the throughput of a parallel configuration {\em and its robustness} under preemptions.
\Cref{fig:comparison}d illustrates how \Sys optimizes liveput. 
After observing two preemptions ($i=2$ in the figure), \Sys anticipates that the cloud has reached its capacity limit and expects additional preemptions in the near future. 
Therefore, instead of maintaining two pipelines each with five instances, which maximizes throughput, \Sys keeps four instances on each pipeline, which is more robust under additional preemptions and maximizes liveput. 
This allows \Sys to cheaply handle future preemptions using lightweight live migrations ($i=3, 4$ in the figure). 

There are three key challenges \Sys must address to optimize liveput: (1) predicting liveput, (2) handling preemptions, and (3) discovering parallel configurations to maximize liveput. We elaborate these challenges and the main ideas \Sys uses to overcome them.

First, spot instances can be preempted and reallocated due to many reasons (e.g., market price changes, resource constraints) at any time.
It is challenging to know ahead of time when and which specific instances will be preempted/allocated by the cloud provider; nor does the cloud provider provides any hints or auxiliary information on how instance preemption and addition decisions are made.
However, estimating the liveput of a parallel configuration requires considering a variety of preemption scenarios.

Instead of predicting preemptions and allocations for individual instances, \Sys uses a two-level approach to forecasting the availability of instances at a coarse granularity. First, the {\em availability predictor} takes the instance preemption and allocation history as input and only predicts the {\em number of available instances} in the near future. Second, the {\em Monte Carlo preemption sampler} uses the predicted instance availability to sample preemptions. This two-level approach allows \Sys to employ a lightweight predictor to forecast spot-instance availability and quickly estimate the liveput of different parallel configurations. 

Second, existing checkpoint- and redundancy-based approaches to handling preemptions introduce significant memory and computation overheads.
Checkpoint-based systems (e.g., Varuna~\cite{athlur2022varuna}) omit all model updates since the last checkpoint after each preemption, and periodically saving and loading checkpoints introduce additional overheads.
These overheads are substantial even by adopting fine-grained checkpointing mechanisms~\cite{mohan2021checkfreq} for better overlapping (see \S\ref{eval:end_to_end}).
Meanwhile, redundancy-based systems (e.g., Bamboo~\cite{bamboo}) require redundant computation on each instance even in the absence of preemptions, which decreases training throughput and increases monetary cost due to redundant computations.

To effectively handle preemptions, \Sys uses a lightweight live migration mechanism that allows DNN training to proceed despite losing instances and without introducing redundant computation as done by prior work. 
To achieve this goal, \Sys's live migration mechanism always uses the same number of samples to update model's parameters in each training iteration and opportunistically reorder samples to avoid redundant computation or restarting training.
This approach preserves model accuracy by leveraging the stochastic nature of DNN training --- all training samples are drawn independently from an intrinsic data distribution and {\em reordering samples does not affect model accuracy}~\cite{bottou2012stochastic}.

Third, optimizing liveput requires reasoning about instance preemptions and allocations and quickly adapting to new resources allocations while minimizing transition cost. 
Recent work (e.g., PipeDream~\cite{pipedream-2bw} and Alpa~\cite{zheng22-alpa}) has proposed a variety of techniques to automatically discover throughput-optimized parallel configurations for DNN training. However, all these approaches assume a fixed set of GPUs and do not apply to spot-instance training.

To address this challenge, \Sys's {\em liveput optimizer} formulates the problem of maximizing liveput as an optimization task and uses a novel dynamic programming algorithm to explore the search space of parallel configurations that combine data and pipeline parallelism and discover an {\em optimal} parallel configuration in the search space.

\begin{figure}
    \centering
    \includegraphics[width=1.0\linewidth]{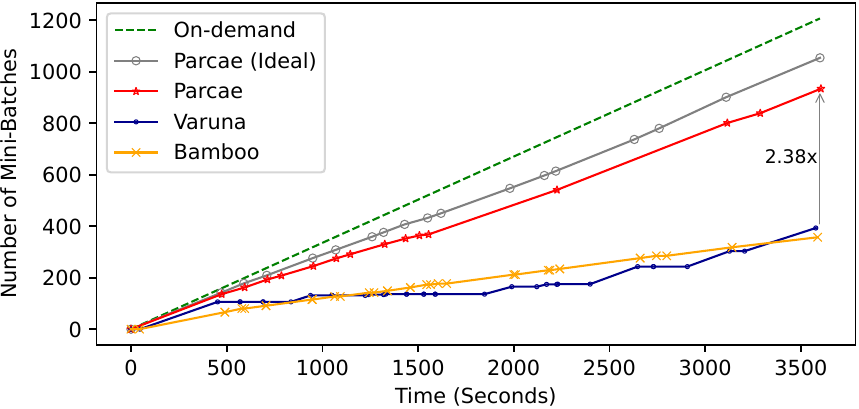}
    \vspace{\captionvspace}
    \caption{Comparing \Sys and prior work for training GPT-2~\cite{gpt2} on 32 spot GPU instances. Note that \Sys, Bamboo, and Varuna use an identical preemption trace.}
    \label{fig:intro_graph}
\end{figure}

The above techniques allow \Sys to significantly outperform prior work. \Cref{fig:intro_graph} compares \Sys against Bamboo and Varuna for training GPT-2 on 32 spot V100 GPU instances on AWS using a collected preemption trace. \Sys outperforms Bamboo and Varuna by 2.4$\times$ under the same preemptions. 
The grey curve shows an {\em ideal} case, where \Sys knows {\em all} future preemptions and allocations and maximizes liveput accordingly. 
\Sys achieves {\em 89\%} efficiency of the ideal case.
We have evaluated \Sys on a variety of DNN models and preemption traces and shown that \Sys outperforms Varuna by up to 9.9$\times$ and Bamboo by up to 10.8$\times$. 
Moreover, our evaluation shows that Bamboo and Varuna cannot scale to large models --- for certain spot-instance traces, both of them cannot make {\em any} progress for training GPT-3~\cite{gpt3} with 6.7 billion parameters, while \Sys can achieve almost identical performance as its ideal case (i.e., knowing all future preemptions and allocations). 

This paper makes the following contributions:
\begin{itemize}
\item We propose liveput, a novel metric that simultaneously consider the performance and robustness of a parallelization strategy for DNN training on spot instances.
\item We build \Sys, a liveput-optimized system for spot-instance training that accurately predicts instance availability, cheaply handles preemptions, and efficiently optimizes training performance under preemptions.
\item We evaluate \Sys and show that it outperforms Varuna and Bamboo by up to 9.9$\times$ and 10.8$\times$, and supports training large-scale models on spot instances.
\end{itemize}
\section{Background}
\label{sec:background}

\subsection{Distributed DNN Training}
\label{bg:dist}

\paragraph{Data parallelism.}
Data parallelism~\cite{pytorch,Tensorflow} is the most widely used parallelization strategy in distributed DNN training. 
Each GPU has a model replica and performs forward and backward computations for different batches of data samples independently. It requires to synchronize model gradients (e.g., All-Reduce~\cite{nccl}) before mode update.

\paragraph{Pipeline parallelism.}
Pipeline parallelism~\cite{gpipe} partitions DNN model into different stages with data dependency. Each stage is trained on one GPU, and different GPUs communicate activations and corresponding gradients, which are computed by forward and backward computation respectively, instead of parameter gradients. A mini-batch of training samples is split into multiple micro-batches in pipeline training and pipeline parallelism exploits the opportunity to parallelize the computations of different micro-batches.

\paragraph{Hybrid data and pipeline parallelism.}
Some studies~\cite{PipeDream, dapple} combine data and pipeline parallelism to further accelerate the training of large models. Given a number of GPUs, the training throughput varies for different parallel configurations, which describes the number of stages and data-parallel pipelines it owns. Some recent systems (e.g., FlexFlow~\cite{unger2022unity}, Alpa~\cite{zheng22-alpa}, Galvatron~\cite{miao2023galvatron}) further involve more complicated model parallelism to benefit distributed training of particular DNNs. 
However, they can not be applied on spot instance with dynamic device membership.
Our approach considers hybrid data and pipeline parallelism, follows Varuna and Bamboo, and leaves the exploration of more fine-grained model parallelism as our future work.

\subsection{Spot-Instance Training}
\label{bg:spot}
Recent frameworks~\cite{torchelastic,athlur2022varuna,bamboo} exploit cheap but preemptible instances provided by clouds to train DNN models on. Torch-Elastic~\cite{torchelastic} focuses on elastic data parallelism training and cannot be adopted to large models, where pipeline parallelism is definitely needed. 
Since the availability of spot instances varies significantly and frequently, it is critical to decide the parallel configuration for a DNN model in response to preemptions and allocations. Bamboo~\cite{bamboo} keeps the pipeline depth fixed and varies the number of pipelines according to the availability of spot instances. This mechanism makes it difficult for Bamboo to utilize spot instances, which have low availability, for large models that require a long pipeline. Varuna~\cite{athlur2022varuna} introduces job morphing to dynamically change the parallel configuration and maximize throughput for a given number of spot instances. For instances with low preemption rate or models with negligible reconfiguration cost, switching to the optimal parallel configuration is definitely optimal. However, the current spot instance market and DNN models violate the two conditions, making it sub-optimal to always adopt the parallel configuration with the optimal throughput.

\section{Liveput}
\label{sec:liveput}
This section introduces {\em liveput}, a new metric for DNN training that describes the {\em expected} training throughput of a parallelization strategy on spot instances by simultaneously considering its throughput and robustness under preemptions.

\subsection{Definition of Liveput}

To address the challenge mentioned above, we introduce liveput, a novel metric for distributed DNN training on spot instances that considers both the performance of a DNN system as well as potential preemptions. 

\begin{definition}[Liveput]
\label{def:liveput}
Let $(D, P)$ denote the parallel configuration of a DNN training job, where $P$ is the number of pipeline stages, and $D$ is the number of data-parallel pipelines. The \textbf{liveput} of this training job is the expectation of its throughput under all possible preemption scenarios:
\begin{align}
    \label{eq:liveput}
    \small
    \LPT(D,P,\mathcal{V})=\mathop{\mathbb{E}}_{\vec{v} \sim \mathcal{V}}[\TPT(D_{\vec{v}}, P_{\vec{v}})]
\end{align}
where $\mathcal{V}: \{0, 1\}^{D \times P}\to [0, 1]$ is the probability distribution of all preemption scenarios. Each $\vec{v}$ is an preemption indicator vector, ${v}_{k}=1$ if instance $k$ will be preempted and ${v}_{k}=0$ otherwise. $\TPT(D_{\vec{v}}, P_{\vec{v}})$ is the throughput of the new parallel configuration $(D_{\vec{v}}, P_{\vec{v}})$ after preemption $\vec{v}$.
\end{definition}

\begin{figure}[t]
    \centering
    \includegraphics[width=1.0\linewidth]{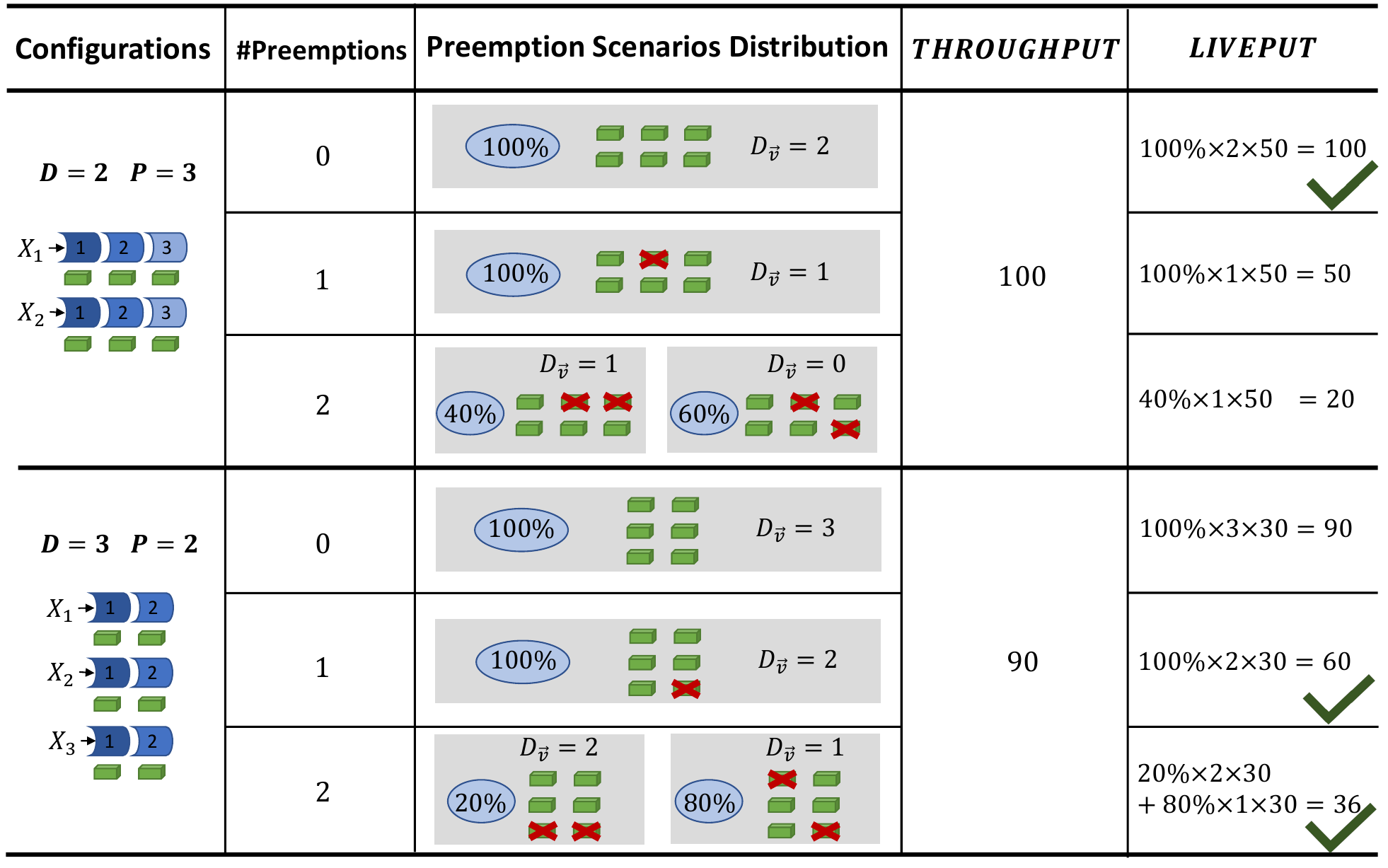}
    \vspace{\captionvspace}
    \caption{Comparing the liveput and throughput of different parallel configurations and preemption scenarios.} 
    \label{fig:liveput}
\end{figure}

Note that we follow prior work~\cite{athlur2022varuna, bamboo} and focus on data- and pipeline-parallel DNN training in this paper, while the liveput definition can easily generalize to other parallel configurations such as model~\cite{FlexFlow} and reduction~\cite{unger2022unity} parallelism.

\subsection{Comparing Liveput and Throughput}
A key advantage of liveput is that it considers how the performance of a parallel configuration changes under different preemption scenarios. 
\Cref{fig:liveput} demonstrates this advantage with a DNN training example on six spot instances with two possible parallel configurations: $\{D=2, P=3\}$ and $\{D=3, P=2\}$. 
For simplicity, we assume the throughput of a pipeline with three (or two) stages is 50 (or 30) samples/second and ignore the parameter synchronization cost.
We compare the two parallel configurations under three preemption scenarios: (a) no preemption, (b) one preemption, and (c) two preemptions. 
We also assume that the preemption probabilities of all instances are the same.

\Cref{fig:liveput} compares the throughput and liveput of the two parallel configurations under the three preemption scenarios. Throughput is independent of instance preemptions; therefore, $\{D=2, P=3\}$ achieves a higher throughput than $\{D=3, P=2\}$ for all cases. On the other hand, liveput considers the amount of possible preemptions as well as the distribution of these preemptions over spot instances.
When there is no preemption (i.e., fixed resource allocation), liveput is equivalent to throughput. 
Once a concrete future preemption scenario is given as a prior condition, the corresponding liveput can be treated as the effective throughput after such a preemption.
For example, under preemption of 1 or 2 instances, the configuration $\{D=3, P=2\}$ achieves higher effective throughput than $\{D=2, P=3\}$.
Intuitively, due to data dependencies between the pipeline stages, longer pipelines are more vulnerable to preemptions, since a single preemption would invalidate an entire pipeline within a mini-batch, and shorter pipelines exhibit better elasticity and resilience under frequent preemptions. 
Existing throughput-optimized approaches fail to consider this trade-off when estimating training efficiency and may make suboptimal decisions.

\section{\Sys Overview}
\begin{figure}
    \centering
    \includegraphics[width=1.0\linewidth]{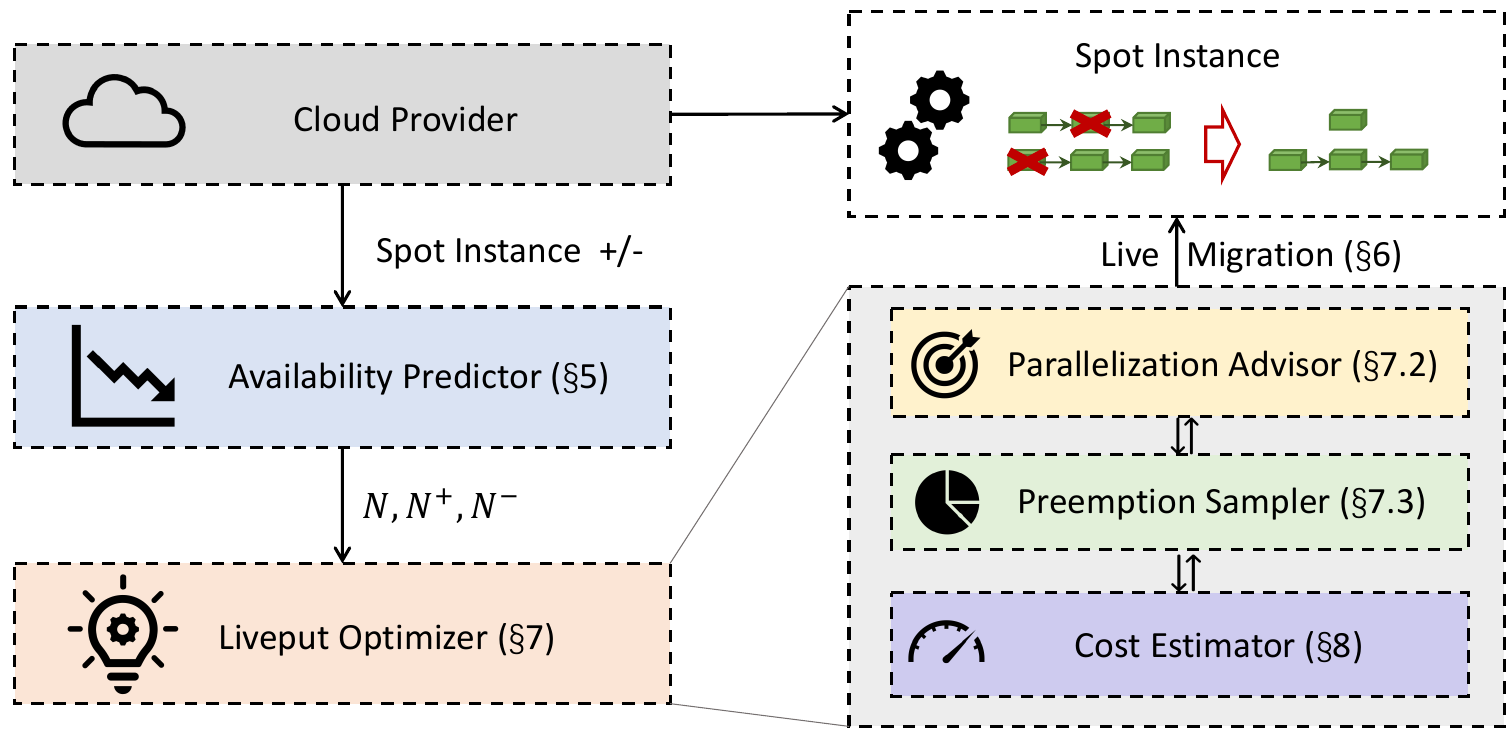}
        \vspace{\captionvspace}
    \caption{An overview of \Sys.}
    \label{fig:overview}
\end{figure}

\Cref{fig:overview} shows an overview of \Sys, a liveput-optimized system for DNN training on spot instances.
Computing liveput requires predicting instance preemptions and allocations. 
Since predicting instance-wise availability is infeasible (\S\ref{sec:unpredict}), \Sys uses a two-level approach to forecasting the availability of all instances at a coarse granularity, where an {\em availability predictor} takes the instance preemption/allocation history as input and only predicts the {\em number of available instances} in the future, and the Monte Carlo {\em preemption sampler} uses the predicted availability to sample preemptions and allocations.

\Sys's {\em liveput optimizer} takes the predicted instance availability as input and discovers a parallel configuration to maximize the liveput of the DNN model. The liveput optimizer formulates the problem of maximizing liveput as an optimization task and uses a dynamic programming algorithm to discover an optimal parallel configuration.

To migrate across different parallel configurations and handle potential preemptions, \Sys uses three \textit{live migration} strategies. 
These migration strategies leverage statistical robustness of DNN training, allow \Sys to significantly reduce migration and preemption overheads compared to existing checkpoint- and redundancy-based systems.

For the rest of this paper, we introduce \Sys's availability predictor in \S\ref{sec:predictor}, live migration strategies in \S\ref{sec:instance_migration}, and liveput optimizer in \S\ref{sec:advisor}. 
\S\ref{sec:fault} describes how \Sys handles exceptional cases where actual preemptions mismatch \Sys's predictions.
Finally, we discuss \Sys's design and implementation on modern clouds in \S\ref{sec:impl} and evaluate its performance in \S\ref{sec:eval}.

\section{Availability Predictor}
\label{sec:predictor}

\subsection{Instance-wise Availability Unpredictability}
\label{sec:unpredict}

There are several factors that affect spot-instance preemptions and allocations, including the types of the instances a user requires and their availability zones, the price of the current spot instance market, and competitions from other users.
Most existing approaches to predicting the availability of spot instances focus on estimating their prices~\cite{harlap2017proteus,harlap2018tributary}, which cannot be used to estimate their lifetime.
Prior work~\cite{mishra2018survey, harlap2018tributary,li2020spottune,yang2021scheduling} has also tried to predict the reliability of spot instances based on historical data collected from cloud providers.
These attempts rely heavily on the cloud behaviour, which varies across cloud providers and availability zones within a cloud.
Moreover, for a new cloud or zone, applying these data-driven approaches before running a job is expensive and time consuming.
As a result, accurately forecasting individual instances' preemptions (i.e., $\vec{v}$ in \Cref{def:liveput}) is impractical since clouds currently do not support specifying preferences on the instance preemption order (i.e., which instances to preempt first), nor do they provide any auxiliary information that can help understand preemption and allocation decisions.

\setcounter{figure}{5}
\begin{figure*}[t]
    \centering
    \includegraphics[width=0.94\linewidth]{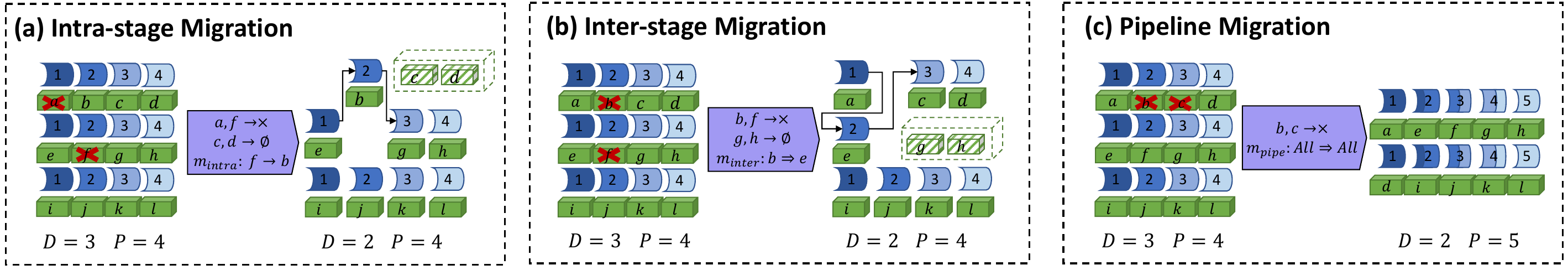}
        \vspace{\captionvspace}
    \caption{Illustrations of different migration strategies over a $3\times 4$ parallel configuration facing 2 preempted instances.}
    \label{fig:migration}
\end{figure*}

\subsection{Statistical Availability Prediction}

To make \Sys a general and practical DNN training system on spot instances, the only visible and reliable information is the past preemption/allocation records of the current user-submitted training job. Instead of forecasting when and which instance will be preempted in the future, \Sys uses a coarse-grained time-series forecasting approach. We observe that it is possible to predict the {\em total amount} of available spot instances for short time intervals in the future and benefit \Sys's proactive optimization performance.

\paragraph{Problem formulation.} We split the timeline of a training job into equally sized intervals, where the length of an interval $T$ is a hyper-parameter. For the $i$-th interval, we define a tupe $(N_i, N_i^+, N_i^-)$ to represent the number of available instances, newly allocated instances, and preempted instances within the $i$-th interval, respectively. We assume that node preemptions and allocations only happen at the beginning of each time interval and that all available spot instances are stable within a time interval; this assumption is reasonable since each interval is small (e.g., 1 minute). Therefore, we have $N_i = N_{i-1} + N^{+}_{i} - N^{-}_{i}$ ($i > 0$).
Instead of predicting $N_i^+$ and $N_i^-$, \Sys's availability predictor only forecasts a sequence of $N_i$ (i.e., overall availability) and uses $N_i$ to derive $N_i^+$ and $N_i^-$.
This design is based on an important observation that a cloud does not preempt existing instances and allocate new instances at the same time, therefore $N^{+}_{i}=\max (0, N_i - N_{i-1})$ and $N^{-}_{i}=\max (0, N_{i-1} - N_{i})$.
Formally, in the time-series forecasting problem, an agent takes the instance availability trace in the past $H$ intervals as input and forecasts the instance availability for the future $I$ time intervals:
\begin{align}
    \label{eq:prediction}
    \small
    (N_{i}, \cdots, N_{i+I-1}) = \textproc{Prediction}(N_{i-H}, \cdots, N_{i-1}).
\end{align}
Note that $(N_{i}, \cdots, N_{i+I-1})$ can be used to derive the predicted instance preemptions and allocations for the next $I$ intervals.

\setcounter{figure}{4}
\begin{figure}
    \centering
    \subfloat[ARIMA vs. other models]{
        \centering
        \includegraphics[scale=0.25]{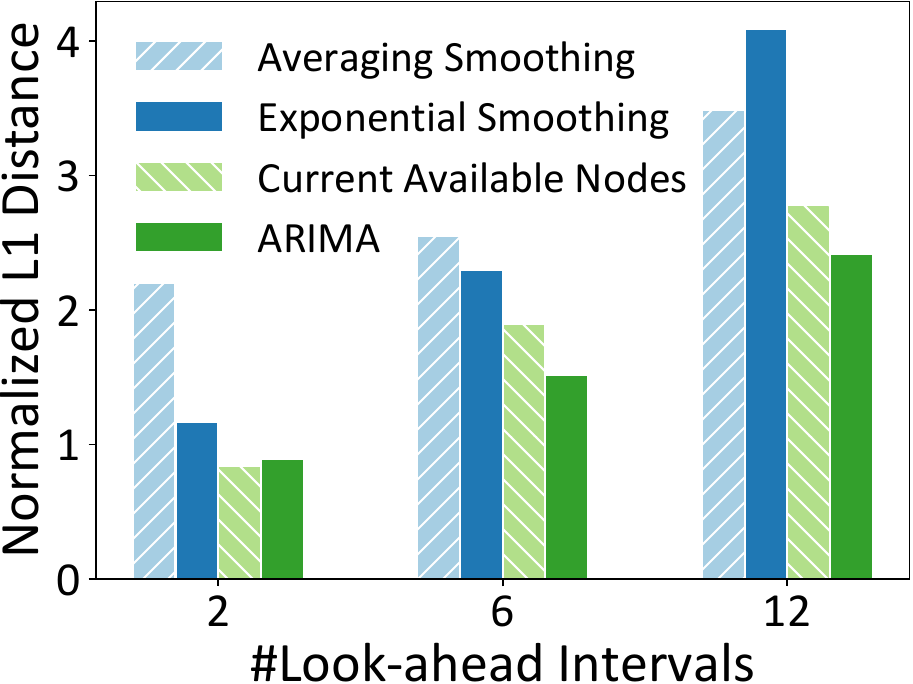}
        \label{fig:arima_vs_others}
    }
    \subfloat[ARIMA-predicted vs. real trace]{
        \centering
        \includegraphics[scale=0.25]{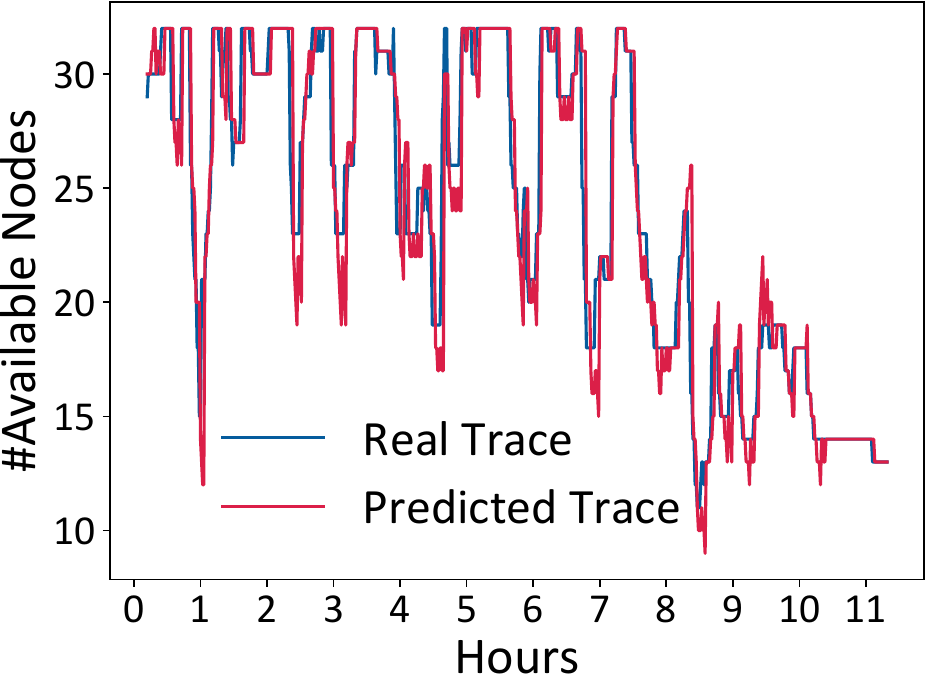}
        \label{fig:arima_vs_real}
    }
        \vspace{\captionvspace}
    \caption{(a) Comparison of normalized L1 distance of predictive performance for ARIMA and other solutions (H$=12$, lower is better). (b) Comparison between ARIMA-predicted trace (H$= 12$, I$= 4$) and the ground truth. }
    \label{fig:predictresult}
\end{figure}
\setcounter{figure}{6}

Limited input data prevents \Sys from using complex prediction models such as deep neural networks. Instead, we propose to leverage lightweight statistical algorithms (e.g., moving averaging, exponential smoothing, current available nodes) and empirically study their performance in \Cref{fig:arima_vs_others} (more details are in \Cref{appendix:arima}). We select the auto-regressive integrated moving average (ARIMA) algorithm~\cite{george1970time} as our availability predictor due to its superior performance. We observe that ARIMA can faithfully describe the tendency of instance availability, as shown in \Cref{fig:arima_vs_real}. Finally, our evaluation on collected real-world preemption/allocation traces further verifies that the ARIMA predictor can help \Sys achieve near-optimal liveput (\S\ref{eval:end_to_end}).

\section{Live Migration}
\label{sec:instance_migration}

This section describes the {\em proactive} migration mechanism of \Sys.
Existing checkpoint- and redundancy-based approaches handle preemptions {\em reactively}, leading to significant overheads.
Instead, we design several fine-grained \textit{live migration} strategies to {\em proactively} handle different future preemption scenarios.
Given the preemption prediction results, \Sys could schedule efficient adjustments in advance to adapt to the dynamic instance availability.

\subsection{Pipeline-aware Preemption Mapping}
\label{subsec:preemption_mapping}

Before introducing live migration, we first discuss the \textit{preemption mapping} step in \Sys. 
Recall that the outputs of the availability predictor (\S\ref{sec:predictor}) only include statistical information (i.e., the number of preemptions or allocations during a time interval).
However, the impact of an instance preemption highly depends on the instance's position in the data- and pipeline-parallel topology.
Therefore, instance-wise preemption predictions (i.e., $\vec{v}$ in \Cref{def:liveput}) is still necessary to make efficient live migration decisions.

To bridge this gap, \Sys uses a probabilistic model to reason about the mapping from preemption events to actual instances.
This preemption mapping is essential for data- and pipeline-parallel training because of the unique dependencies between instances. In particular, instances in the same pipeline have sequential dependencies for both forward and backward computation, and instances in the same stage have synchronization dependencies for parameter synchronization. 
For each preemption event, \Sys assumes that all spot instances may be preempted with the same probability (see the example in \Cref{fig:liveput}).
Note that such an assumption can be replaced by more accurate estimations when additional preemption information is provided by cloud providers.

\subsection{Migration Strategies}
\Sys uses three migration strategies (\Cref{fig:migration}) to handle preemptions: {\em intra-stage}, {\em inter-stage}, and {\em pipeline} migration.

\paragraph{Intra-stage migration.} In pipeline-parallel training, instances in the same stage maintain the same shard of model parameters. 
Therefore, when an instance is preempted, \Sys can opportunistically divert an available instance from the same stage in another broken pipeline. This intra-stage migration allows \Sys to re-establish a complete pipeline.
As shown in \Cref{fig:migration} (a), when instances $a$ and $f$ are preempted, \Sys can replace $f$ by moving $b$ to the second pipeline (e.g., $f\to b$), resulting in two complete pipelines. Intra-stage migration only requires updating the communication routing (e.g., $\to$) of a few instances and does not involve transferring parameters since $b$ and the preempted instance $f$ share the same model parameters and states.

\paragraph{Inter-stage migration.}
When intra-stage migration does not help recover broken pipelines, \Sys opportunistically performs inter-stage migrations, which moves intstances across stages. \Cref{fig:migration} (b) shows an inter-stage migration, where instances $b$ and $f$ are preempted, and \Sys moves $e$ from the first stage to the second stage of the first pipeline (e.g., $b\Rightarrow e$), resulting in two complete pipelines. Inter-stage migration requires transferring model parameters (e.g., $\Rightarrow$) as the instances keep the model parameters and states of different stages. Both intra- and inter-stage migrations preserve pipeline depth and manage to recover as many data-parallel pipelines as possible.

\paragraph{Pipeline migration.}
Changing the pipeline depth is an important choice for maximizing training efficiency. Compared with the other two migration strategies, pipeline migration requires repartitioning the DNN model into a different number of pipeline stages, which involves significant migration overheads as instances need to broadcast their model parameters (e.g., $All\Rightarrow All$). Pipeline migration is similar to the reconfiguration mechanism in prior work (e.g., Varuna~\cite{athlur2022varuna}, Bamboo~\cite{bamboo}) to handle instance preemptions.

\Sys makes migration decisions by considering the current parallel configuration, the new optimized parallel configuration and the actual preemptions. Given the probabilistic mapping of predicted preemptions, \Sys automatically renewals the optimal parallel configuration and the migration strategy (\S\ref{subsec:st_opt}). Once the prediction mismatches with the actual availability, \Sys adjusts the parallel configuration as well as the corresponding migration strategies for adaptation (\S\ref{sec:fault}). The actual migration decisions are finalized when preemptions really happen, and \Sys leverages the grace period (e.g., 30s on Azure~\cite{azure_spot}) to perform these migrations.

\section{Liveput Optimizer}
\label{sec:advisor}
This section describes \Sys's {\em liveput optimizer}, which determines the parallel configurations of training a DNN model on spot instances to maximize its liveput.

\subsection{Problem Definition}
We formulate liveput maximization as an optimization problem, where the objective is to discover a sequence of parallel configurations to maximize the committed training samples in expectation of spot instance availability. 
The sequence length is set to be consistent with the number of time intervals predicted by the availability predictor (\Cref{sec:predictor}). 
Formally, the objective function $\Phi$ is the accumulated number of committed training samples during the $I$ time intervals:
\begin{align}  
    \small
    \Phi(\mathbf{D},\mathbf{P} \mid \mathbf{N}) = \sum_{i=0}^{I-1}\phi(D_{i}, P_{i}, N_{i} \mid D_{i+1}, P_{i+1}, N_{i+1}),
\end{align}
where $N_i$ is the predicted number of available instances (see \Cref{sec:predictor}) at the $i$-th time interval.
Recall that \Sys derives $N_{i+1}^{-}$ (i.e., the number of instances to be preempted) and $N_{i+1}^{+}$ (i.e., the number of instances to be launched) from $N_{i}$ and $N_{i+1}$.
In addition, the preemption distribution $\vec{v}_{i+1}$ (\Cref{def:liveput}) is generated from $N_i$ and $N_{i+1}^{-}$ using the probabilistic preemption model developed in \Cref{subsec:preemption_mapping}.
Finally $\phi$ calculates the number of committed samples within a interval:

\vspace{-2mm}
\begin{align}  
    \small
    \label{eq:phi}
    \phi(& D_{i},  P_{i}, N_{i} \mid D_{i+1}, P_{i+1}, N_{i+1}) \\ \nonumber 
    & = \mathop{\mathbb{E}}_{\vec{v}_{i+1}}[\LPT(D_{i+1},P_{i+1} \mid \vec{v}_{i+1})\times T_{\text{eff}}],\\  
    T_{\text{eff}} & = T - T_{\text{mig}}(D_i, P_i,  D_{i+1}, P_{i+1} \mid \vec{v}_{i+1}), \nonumber
\end{align}

\noindent where $T$ and $T_{\text{eff}}$ are the length of the time interval and effective training time after migrations, respectively, 
and $T_{\text{mig}}$ is the migration overhead. 
Note that $\phi$ extends liveput by making the preemption distribution $\vec{v}_{i+1}$ a prior. With these definitions, the objective of the liveput optimizer is:

\vspace{-2mm}
\begin{align}  
    \label{eq:target}
    \small
    \argmax_{\mathbf{D}, \mathbf{P}} \Phi(\mathbf{D}, \mathbf{P} \mid \mathbf{N})
\end{align}
where $\mathbf{N}=\{N_1, N_{2}, \cdots, N_I\}$  is the output of the availability predictor, and \Sys discovers a sequence of parallel configurations $(\mathbf{D}, \mathbf{P})$ to maximize liveput. 

\subsection{Parallelization Advisor}
\label{subsec:st_opt}

\Sys uses a dynamic programming algorithm to explore the optimization space and discovers an {\em optimal} sequence of parallel configurations. 
Specifically, let $F(i+1, D_{i+1}, P_{i+1})$ represent the maximal number of committed training samples at the end of the $i$-th time interval, which uses parallel configuration ($D_{i+1}, P_{i+1}$).
We start from $F(0, D_{0}, P_{0})=0$ and have the following optimal substrates:

\vspace{-2mm}
\begin{align}
    \small
    \label{eq:opt_dp}
    & F(i+1, D_{i+1}, P_{i+1}) \\ \nonumber 
    = & \max_{\forall D_{i}\times P_{i}\leq N_i}\left\{
    \begin{aligned} 
    & F(i, D_{i}, P_{i}) + \\
    \phi(D_{i}, & P_{i}, N_{i} \mid D_{i+1}, P_{i+1}, N_{i+1})
    \end{aligned}
    \right\},
\end{align}

\noindent and figure out the final target as $\max_{\forall D_{I}\times P_{I}\leq N_I} \left\{F(I, D_{I}, P_{I})\right\}$.

The DP algorithm considers all possible parallel configurations that satisfy resource constraints (i.e., $D_{i}\times P_{i}\leq N_i$), and $\phi(D_{i},  P_{i}, N_{i} \mid D_{i+1}, P_{i+1}, N_{i+1})$ is the product of two terms in \Cref{eq:phi}. 
Here the exploration adapts a similar search space as Varuna with a size of $\mathcal{O}(N\log N)$, which is large enough for most recent large DNNs consisting of a stack of homogeneous layers. It is also possible to extend to a larger search space (e.g., Alpa) for more complicated workloads.
The first term $\LPT$ can be replaced by $\TPT(D_{i+1},P_{i+1})$, where $(D_{i+1},P_{i+1})$ is the new parallel configuration after live migration. 
Note that $(D_{i+1},P_{i+1})$ should be a feasible model partition that satisfies the device memory capacity.
For unfeasible cases that violate memory constraints, their $\TPT$ is set to be zero.

The second term $T_{\text{eff}}$ depends on the preemption distribution, $(D_{i},P_{i})$,  $(D_{i+1},P_{i+1})$, and the migration strategy to transit from $(D_{i},P_{i})$ to $(D_{i+1},P_{i+1})$.
Given a pair of parallel configurations $(D_{i},P_{i})$ and $(D_{i+1},P_{i+1})$, there may exist multiple migration strategies with different overheads $T_{\text{mig}}$, and the cost of each migration strategy also depends on the DNN workload.
\Sys uses a {\em cost estimator} (\Cref{sec:cost}) to estimate $T_{\text{mig}}$ for different migration strategies.
If the pipeline depth changes (i.e., $P_{i+1} \neq P_{i}$), \Sys infers that pipeline migration is performed. Otherwise, $T_{\text{mig}}$ should be attributed to either inter- or intra-stage migrations. When both of them are applicable, \Sys selects the one with lower migration cost.
In the absence of preemptions (i.e., $N_{i+1}=N_{i}$), there can be no migration cost if $(D_{i+1}, P_{i+1})$ equals to $(D_{i}, P_{i})$.

\subsection{Preemption Mapping Sampler}
\label{subsec:preemption_sampler}
As introduced in \Cref{subsec:preemption_mapping}, preemption mapping is necessary to reason about live migration, since preemptions at different positions in the data- and pipeline-parallel topology require different migration strategies.
Given $N_{i}$ spot instances, among which $N^{-}_{i+1}$ are to be preempted, the number of possible preemption mappings on a $D\times P$ topology grows exponentially in $N^{-}_{i+1}$.
The large preemption mapping space makes it infeasible to explicitly consider all preemption scenarios or analyze the exact solutions mathematically.

To address this issue, \Sys uses sampling techniques to explore the mapping space and quickly discovers reasonable accurate approximations.
Specifically, \Sys applies Monte Carlo (MC) sampling over the large space of all preemption scenarios and randomly samples $\vec{v}$ while preserving $N^{-}_{i+1}=\sum_{j=1}^{N_i} v_j$. 
For each sampled $\vec{v}$, \Sys identifies the corresponding migration costs. 
\Sys ensembles multiple trails of sampling to approximate the expectation in \Cref{eq:phi}.
Note that this sampling step can be done offline in advance, therefore it does not block the dynamic programming optimization procedure.
This allows parallelization advisor to quickly compute new parallel configurations and migration strategies during spot-instance training.

\section{Exception Handling}
\label{sec:fault}

This section describes how \Sys handles exceptional cases where actual spot-instance preemptions mismatch \Sys's predictions or the suggested parallel configuration is not compatible with the available spot instances.

\paragraph{Parallelization adaptation.}
Compared to prior work, \Sys proactively adjusts parallel configurations by predicting instances' availability and planning live migrations ahead. However, if actual preemptions rarely differ from predictions, the liveput optimizer may not work on available spot instances. To address this issue, \Sys includes a \textit{configuration adaptation} step to adjust the target parallel configuration before live migration. 
Specifically, when the number of actual available spot instances is greater (or less) than the predicted $N_i$, \Sys adds (or drops) data-parallel pipelines while preserving the pipeline depth. When available spot instances cannot even formulate a single pipeline, it will try to re-partition the pipeline into fewer stages.
This adaptation ensures a feasible configuration without significant migration overheads, performing at least as well as existing throughput-optimized approaches that reactively handle preemptions when predictions go wrong.

\paragraph{Fault tolerance.}
Even if the predictions align well with actual preemptions, there still exist rare cases where the migration strategies do not work. For example, if all instances in one stage are preempted, both inter- and intra-stage migration cannot recover this stage's status. \Sys uses a cheap, in-memory checkpointing mechanism (\S\ref{impl:cpu_ps}) to handle these cases. In addition, for the extreme cases where the number of available instances is less than the minimum feasible pipeline depth $P$, the training process has to be suspended until new spot instances are available.

\section{\Sys's Design and Implementation}
\label{sec:impl}

\begin{figure}[t]
    \centering
    \includegraphics[width=0.9\linewidth]{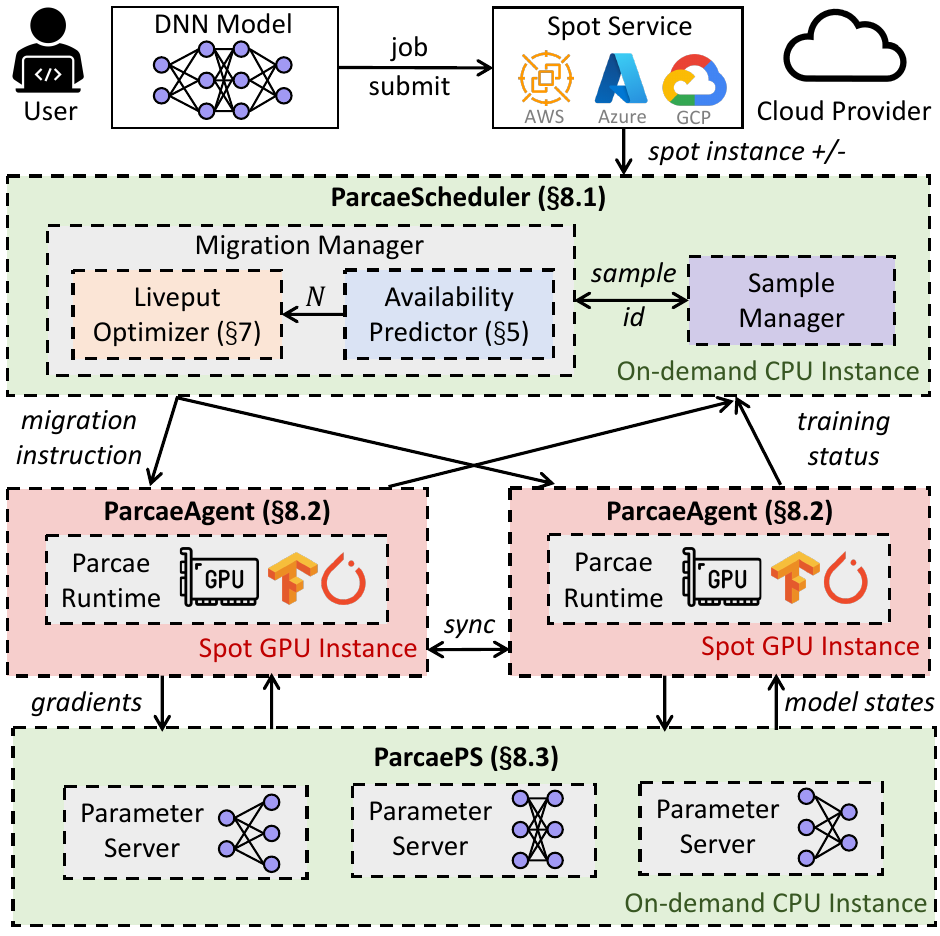}
       \vspace{\captionvspace}
    \caption{Overview of \Sys's design and implementation.}
    \label{fig:impl}
\end{figure}

\Sys consists of three main components as illustrated in \Cref{fig:impl}.
First, \SysSched (\S\ref{sec:agent}) runs persistently on one on-demand CPU instance, determining the migration schedule based on our liveput optimizer and availability predictor. It also manages the training data samples to maintain the training semantics.
Second, each spot GPU instance runs a \SysAgent (\S\ref{sec:agent}), which performs assigned training workload, monitors training progress, and executes the migration strategies issued by the \SysSched.
Third, \SysPS (\S\ref{impl:cpu_ps}) runs on several on-demand CPU instances to keep model checkpoints for rare rollback cases.

\Sys's implementation consists of $\sim8K$ LoC in Python and takes PyTorch~\cite{li13pytorch} as the default runtime. Communications between \SysSched and \SysAgent use \texttt{etcd}~\cite{etcd}, a distributed key-value store. 
We implement live migration strategies by modifying DeepSpeed~\cite{rasley2020deepspeed}. 
We show the workflow of \SysSched and \SysAgent in Algorithm~\ref{algo:sched} and introduce the detailed components as follows.

\subsection{\SysSched}
\label{sec:sched}

\SysSched has two major components: a {\em migration manager} and a {\em sample manager}.
The former is responsible for parallelization and live migration, and the latter handles the data samples distribution.

\paragraph{Migration manager.} 
As shown in Algorithm~\ref{algo:sched}, the migration manager keeps receiving instance availability information (i.e., preemption or allocation interruptions) from the cloud provider and updating the current number of available instances (line 3).
As discussed in \S\ref{sec:fault}, the parallel configuration $(D_i, P_i)$ computed in the previous iteration using predicted availability may be incompatible with the current instances' availability.
To handle this exception, \SysSched first adjusts the target parallel configuration (line 4) and then generates the required migration strategy $S_i$ based on the current and target configurations $(D_{i-1}, P_{i-1})$ and $(D_{i}, P_{i})$. Note that the adaptation step (line 4) is performed before generating the migration strategy (line 5) so it will not involve re-adjustment overheads.
Next, the availability predictor will forecast the number of available instances for a series of future intervals (i.e., $\textbf{N}_{i+1},\cdots,\textbf{N}_{i+I}$) based on the historical information (line 7). 
Finally, the liveput optimizer makes parallelization suggestions for the following time interval using the prediction (line 8).
The workflow continues until the training job is completed.

The handling of instance preemption and allocation interruptions are slightly different. 
Allocations are {\em controllable} as they only occur after we consciously send requests to the cloud, although they may not always succeed.
We let a new instance join after its \SysAgent is successfully initialized.
In contrast, preemptions are {\em passive} and may interrupt instances at any time, which requires additional mechanisms to handle various exceptions.
Fortunately, the clouds usually provide a small grace period to inform the preemption before it happens.
As the duration is usually enough to finish a mini-batch's training, we utilize the preemption notice to simplify the implementation and enforce instances to be preempted only at the mini-batches' boundaries.
\Sys also handles rare failures that may interrupt training process, in which case \SysSched restarts training using the latest checkpoint in \SysPS, avoiding losing model updates.

\begin{algorithm}[t]
\small
\begin{algorithmic}[1] 

\LeftComment{\SysSched}
\Function{MigrationManager}{$D_0,P_0$}
    \For{$i$ in $1$, $2$, $3$, $\cdots$}
        \State  $N_{i}$ $\gets$ Receive availability info from cloud provider
        \State $(D_i, P_i) \gets$ \texttt{AdjustParallelConfiguration}($N_i$)
        \State $S_i \gets$ \texttt{GetMigrationStrategy} ($(D_{i-1}, P_{i-1})$, $(D_i, P_i)$)
        \State Send migration strategy $S_i$ to all \SysAgents
        \State $N_{i+1},\cdots,N_{i+I}$ $\gets$ \texttt{AvailPredictor}($N_{i-H+1}, ..., N_{i}$)
        \State $(D_{i+1}, P_{i+1}) \gets $ \texttt{LiveputOpt}\big($(D_i, P_i), N_i, ..., N_{i+I}$\big) 
        \If{job completes}
        \State \textbf{break}
        \EndIf
    \EndFor
\EndFunction

\LeftComment{\SysAgent}
\Function{ParcaeRuntime}{model, batch\_size}
\While{job does not complete}
    \State Receive migration instruction $m$ from \SysSched
    \State Apply migration instruction $m$ if $m$ is not empty
    \State $X, Y \gets$ \texttt{DataLoader}(batch\_size)
    \State \texttt{Train}(model, $X$, $Y$)
\EndWhile
\EndFunction

\end{algorithmic}
\caption{Workflow of \Sys components.}
\label{algo:sched}
\end{algorithm}

\paragraph{Sample manager.}
The training dataset is divided into mini-batches of fixed size and trained by DNNs iteratively.
Each mini-batch of samples are ``committed'' after each iteration.
However, preemptions may terminate training at any time, resulting in uncommitted mini-batches (\Cref{fig:comparison}). 
To guarantee the same training semantics as on-demand instances, the sample manager tracks each data sample, records all uncommitted samples' indices, and makes them rejoin the training process later.
This guarantees that all data samples are trained exactly once per epoch, preserving identical theoretical convergence property as the original data feeding order. We also provide a convergence experiment in \Cref{fig:convergence} to verify its training correctness.

\subsection{\SysAgent}
\label{sec:agent}

A \SysAgent runs on each spot GPU instance to interact with \SysSched as shown in  \Cref{algo:sched}. 
It repeatedly receives a migration instruction from the \SysSched (line 13).
If no migration is required, the \SysAgent requests a batch of training samples and starts model training (line 15-16).
Otherwise, it performs the assigned migration instruction (line 14).
\SysAgent manages to reuse the current model states to alleviate checkpoint overheads and rollbacks.
For example, intra-stage migration is implemented by rebuilding communication groups and reusing previous model states on each GPU.
For inter-stage and pipeline migration, additional costs are required for loading the latest model states from other instances via GPUs' peer-to-peer communications.
Specially, if all instances of a stage are preempted, all the \SysAgents have to roll back to a previous checkpoint.
In this way, \SysSched automatically generates the most efficient migration strategy and let the \SysAgents transit to the target parallel configuration.
Note that, the \SysSched also notifies a \SysAgent if it will be preempted or stay idle (i.e., $N_i-D_i\times P_i$ instances will be idle) by sending a halt or termination instruction to the \SysAgent.

\subsection{\SysPS}
\label{impl:cpu_ps}
\Sys needs checkpoints to handle rare cases as introduced in \S\ref{sec:fault}.
Unlike prior checkpointing approaches relying on expensive cloud storage (e.g., S3 on AWS), \Sys employs several cheap on-demand CPU instances (e.g., {\tt c5.4xlarge} instance, $0.68\$/hour$) to maintain the latest model states in their DRAM.
Instead of directly communicating model states and weights as prior checkpointing approaches,  the \SysPS maintains an up-to-date checkpoint by iteratively synchronizing gradients with spot GPU instances to update the model states the on CPU side (e.g., parameters and optimizer states), which reduces communication by 5$\times$ for stateful optimizers (e.g., Adam~\cite{kingma2014adam}) in the FP16 format~\cite{ren2021zero}.
\Sys also partitions gradients into small pieces for better overlapping and prevents bandwidth competition with cross-stage activation transfer.

\subsection{Cost Estimator}
\label{sec:cost}

We develop a {\em cost estimator} to estimate migration cost by considering different preemption scenarios and parallel configurations.
We conduct an empirical study to profile the migration cost and find that it varies across several factors (details in \Cref{appendix:mig_cost}).
Some of these terms have relatively fixed overheads like CUDA context initialization (less than 10s).
The communication group updating and model building costs are associated with the parallel configuration (less than 30s).
The model transfer cost varies considerably according to the preemptions (up to 60s).
We consider the instance network topology for each preemption scenario and adopt an $\alpha-\beta$ model~\cite{valiant1990bridging} to accurately estimate the communication cost.

\section{Evaluation}
\label{sec:eval}

\begin{table}[t]
\centering
\caption{Overview of the four trace segments evaluated.}
\label{eval:traces}
\small
\begin{tabular}{ l|llll }
\hline
Trace & $H_{A}D_{P}$ & $H_{A}S_{P}$ & $L_{A}D_{P}$ & $L_{A}S_{P}$ \\
\hline
\textit{Availability} & High  & High & Low & Low \\
\textit{Preemption intensity} & Dense & Sparse & Dense & Sparse \\
\#avg instances & 27.05 & 29.63 & 16.82 & 14.60\\
\#preemption events & 9 & 6 & 8 & 3 \\
\#allocation events & 8 & 5 & 12 & 0 \\
length & 1h & 1h & 1h & 1h \\
\hline
\end{tabular}
\\[5pt]

\includegraphics[scale=0.35]{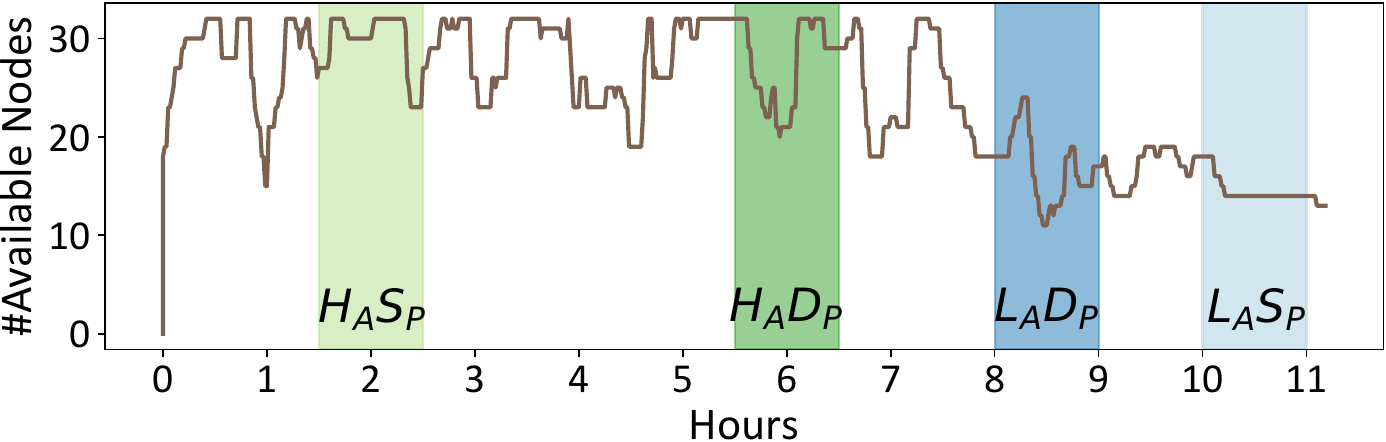}
    \vspace{\captionvspace}
\captionof{figure}{The complete trace and segments of four scenarios.}
\label{fig:trace}
\end{table}

\begin{figure*}
    \centering
    \subfloat[\Sys throughput.]{
        \centering
        \includegraphics[width=0.7\linewidth]{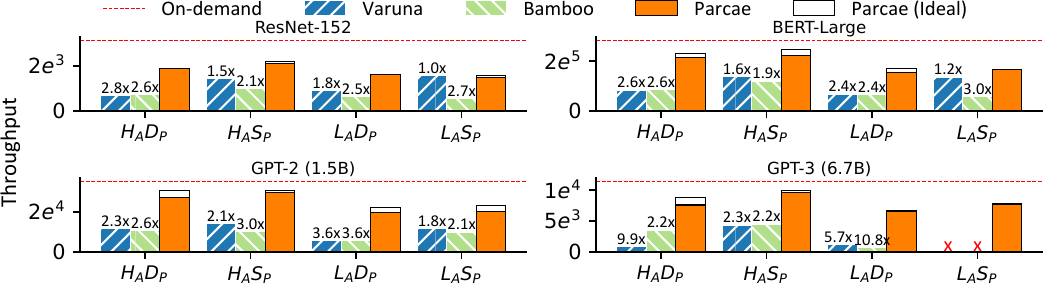}
        \label{fig:end_to_end}
    }
    \subfloat[The effect of $I$ in predictor.]{
        \centering
        \includegraphics[width=0.25\linewidth]{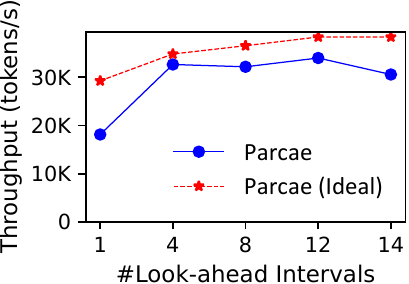}
        \label{fig:ahead_ablation}
    } 
    \vspace{\captionvspace}
    \caption{(a) Training throughput comparison among existing frameworks and \Sys on four traces. The dotted on-demand line shows the best throughput with on-demand instances. The numbers over the bars represent the speedup of \Sys over Varuna and Bamboo respectively. (b) The GPT-2 training throughput for the $H_AD_P$ trace with different look-ahead intervals.}
\end{figure*}

\subsection{Experimental Setup}
\paragraph{DNNs.} 
We select five popular DNNs for various applications.
ResNet-152~\cite{resnet} and VGG-19~\cite{VGG} are CV tasks, and we use CIFAR-100~\cite{cifar100} as the training dataset.
BERT~\cite{bert}, GPT-2~\cite{gpt2}, and GPT-3~\cite{gpt3} are popular model architectures for NLP tasks, and we evaluate them on WikiText-2~\cite{wikitext2}. 
We use GPT-2 and GPT-3 including 1.5 and 6.7 billion model parameters respectively. More setting details are in \Cref{appendix:exp_detail}.

\paragraph{Traces.} 
Due to the dynamic availability of spot instance, it is almost impossible to evaluate different systems on real spot instances multiple times and expect consistent dynamic environments. Instead, to make a fair comparison, we take the real spot instance availability traces and replay them on regular instances. Specifically, we collect a 12-hour trace on a 32-instance cluster with {\tt p3.2xlarge} instances on AWS. Inspired by Bamboo~\cite{bamboo}, we extract representative segments from the whole trace for our evaluation. We design two new measurements for each segment, including the \textit{availability} (i.e., the average number of instances) and the \textit{preemption intensity} (i.e., the number of instance preemption and allocation events). 
Table~\ref{eval:traces} and Figure~\ref{fig:trace} show four extracted 1-hour trace segments based on different availability and preemption intensity.
Traces with over $70\%$ available instances are high availability (i.e., $H_{A}$) traces, otherwise have low availability (i.e., $L_{A}$).
Dense preemption intensity traces (i.e., $D_{P}$) have around 20 instance preemption and allocation events, but sparse preemption intensity traces (i.e., $S_{P}$) only have few. We replay these four trace segments on 32 on-demand V100-16GB GPU instances to simulate spot-instance clusters.

\begin{table}[t]
\centering
\small
\caption{Comparison of monetary cost ($\times1e^{-6}$USD) for different models and approaches. We report per-image cost for ResNet and VGG and per-token cost for BERT and GPT.}
\label{eval:cost}
\scalebox{0.8}{
\begin{tabular}{ cc|cccc }
\hline
Model & Trace & On-Demand & Varuna & Bamboo & \Sys \\
\hline
\multirow{4}{*}{ResNet} & $H_{A}D_{P}$ &  8.68 (2.3$\times$) & 10.86 (2.8$\times$) &  9.77 ( 2.6$\times$) &  3.81 (1$\times$) \\
 & $H_{A}S_{P}$ &  8.68 (2.4$\times$) &  5.32 (1.5$\times$) &  7.61 ( 2.1$\times$) &  3.62 (1$\times$) \\
 & $L_{A}D_{P}$ &  8.68 (3.2$\times$) &  4.89 (1.8$\times$) &  6.72 ( 2.5$\times$) &  2.71 (1$\times$) \\
 & $L_{A}S_{P}$ &  8.68 (3.4$\times$) &  2.43 (1.0$\times$) &  6.96 ( 2.7$\times$) &  2.54 (1$\times$) \\
\hline
\multirow{4}{*}{VGG} & $H_{A}D_{P}$ & 12.43 (2.7$\times$) & 12.10 (2.6$\times$) & 12.11 ( 2.6$\times$) &  4.62 (1$\times$) \\
 & $H_{A}S_{P}$ & 12.43 (2.7$\times$) &  6.52 (1.4$\times$) & 13.12 ( 2.8$\times$) &  4.66 (1$\times$) \\
 & $L_{A}D_{P}$ & 12.43 (3.4$\times$) &  5.43 (1.5$\times$) &  9.40 ( 2.6$\times$) &  3.66 (1$\times$) \\
 & $L_{A}S_{P}$ & 12.43 (4.0$\times$) &  3.37 (1.1$\times$) &  8.88 ( 2.9$\times$) &  3.11 (1$\times$) \\
\hline
\multirow{4}{*}{BERT} & $H_{A}D_{P}$ &  0.10 (2.9$\times$) &  0.09 (2.6$\times$) &  0.09 ( 2.6$\times$) &  0.03 (1$\times$) \\
 & $H_{A}S_{P}$ &  0.10 (2.8$\times$) &  0.06 (1.6$\times$) &  0.06 ( 1.9$\times$) &  0.03 (1$\times$) \\
 & $L_{A}D_{P}$ &  0.10 (3.4$\times$) &  0.07 (2.4$\times$) &  0.07 ( 2.4$\times$) &  0.03 (1$\times$) \\
 & $L_{A}S_{P}$ &  0.10 (4.2$\times$) &  0.03 (1.2$\times$) &  0.07 ( 3.0$\times$) &  0.02 (1$\times$) \\
\hline
\multirow{4}{*}{GPT-2} & $H_{A}D_{P}$ &  0.62 (2.9$\times$) &  0.49 (2.3$\times$) &  0.55 ( 2.6$\times$) &  0.21 (1$\times$) \\
 & $H_{A}S_{P}$ &  0.62 (3.0$\times$) &  0.44 (2.1$\times$) &  0.62 ( 3.0$\times$) &  0.21 (1$\times$) \\
 & $L_{A}D_{P}$ &  0.62 (3.5$\times$) &  0.63 (3.6$\times$) &  0.64 ( 3.6$\times$) &  0.18 (1$\times$) \\
 & $L_{A}S_{P}$ &  0.62 (4.1$\times$) &  0.27 (1.8$\times$) &  0.31 ( 2.1$\times$) &  0.15 (1$\times$) \\
\hline
\multirow{4}{*}{GPT-3} & $H_{A}D_{P}$ &  2.39 (2.5$\times$) &  9.35 (9.9$\times$) &  2.07 ( 2.2$\times$) &  0.94 (1$\times$) \\
 & $H_{A}S_{P}$ &  2.39 (3.0$\times$) &  1.81 (2.3$\times$) &  1.74 ( 2.2$\times$) &  0.80 (1$\times$) \\
 & $L_{A}D_{P}$ &  2.39 (3.6$\times$) &  3.81 (5.7$\times$) &  7.28 (10.8$\times$) &  0.67 (1$\times$) \\
 & $L_{A}S_{P}$ &  2.39 (4.8$\times$) & - & - &  0.49 (1$\times$) \\
\hline
\end{tabular}
}
\end{table}

\subsection{End-to-End Evaluation}
\label{eval:end_to_end}
We first compare the end-to-end training performance between \Sys and existing SOTA spot-instance training systems including Bamboo~\cite{bamboo} (redundancy-based) and Varuna~\cite{athlur2022varuna} (checkpoint-based). We also compare with on-demand instances training approach. The results are displayed in Figure~\ref{fig:end_to_end} and Table~\ref{eval:cost}. In all experiments, \Sys looks ahead 12 intervals based on the availability predictor, while \Sys (Ideal) looks ahead 12 intervals based on truth traces.

\Sys significantly outperforms both Bamboo and Varuna in terms of throughput for almost all the models and preemption traces. 
On average, \Sys delivers an overall of $2.59\times$ higher throughput than Varuna and $3.0\times$ than Bamboo. 
Apparently, \Sys is much more economical than Varuna and Bamboo as it completes more samples with the same monetary costs.
Compared with on-demand instances, \Sys is $3.24\times$ cheaper, and \Sys (Ideal) even achieves competitive throughput, e.g., only $14.2\%$ lower for GPT-2 on high availability traces.
The results also show that the performance of \Sys is quite close (i.e., up to $13.3\%$) to \Sys (ideal).

The performance improvement mainly comes from two aspects. First, \Sys's \liveput optimized configurations balance the trade-off between throughput and available duration, instead of greedily doing expensive reconfiguration like Varuna. While Bamboo maintains a fixed long pipeline depth (e.g., 16 for GPT-2), leading to many unutilized instances, especially for low availability traces.
Second, the migration mechanism in \Sys is highly efficient to handle preemptions. 
Varuna is designed for low preemption environments and relies on shared storage (e.g. S3) to save and load checkpoints.
Although Varuna overlaps checkpoint saving with training iterations, when preemptions happen, it requires rolling back to the last checkpoint and loses tens of seconds’ (i.e., the duration of one complete checkpointing) training progress for large models. To recover from preemptions, Varuna needs to load the last checkpoint from persistent storage and restart training, which is also expensive.
Bamboo is designed for high preemption environments based on redundant computation. It can efficiently handle preemptions, but the redundant computation is inefficient and brings additional synchronization overheads between redundant and normal modules.

\begin{figure}[t]
\centering
\begin{minipage}{.54\linewidth}
    \centering
    \includegraphics[width=\linewidth]{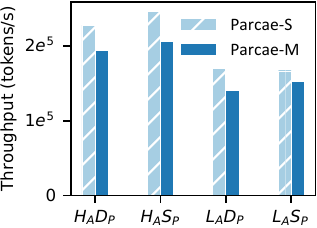}
\end{minipage}%
\hfill
\begin{minipage}{.4\linewidth}
\centering
\scalebox{0.75}{
\begin{tabular}{ ccc }
  \toprule
  & \multicolumn{2}{c}{Monetary Cost} \\[3pt]
  \small
  Trace & \Sys-S & \Sys-M \\[3pt]
  \midrule
  $H_{A}D_{P}$ &  \textbf{3.13} & 3.78 \\[3pt]
  $H_{A}S_{P}$ &  \textbf{3.12} & 3.71 \\[3pt]
  $L_{A}D_{P}$ &  \textbf{2.57} & 3.28 \\[3pt]
  $L_{A}S_{P}$ &  \textbf{2.27} & 3.00 \\
  \bottomrule
\end{tabular}
}
\end{minipage}
    \vspace{\captionvspace}
\caption{The comparison of throughput and monetary cost ($\times 1e^{-8}$USD/token) of BERT for \Sys on single-GPU instances (\Sys-S) and multi-GPU instances (\Sys-M).}
\label{fig:multi_gpu}
\end{figure}

\begin{figure}[t]
    \centering
    \includegraphics[width=1\linewidth]{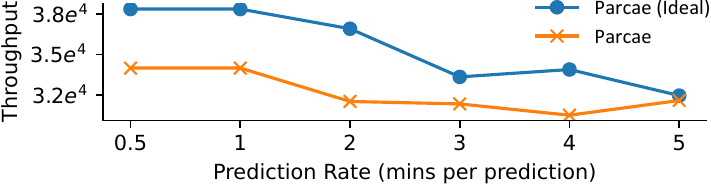}
        \vspace{\captionvspace}
    \caption{GPT-2 training throughput (tokens/s) using the $H_AD_P$ trace with different prediction rates.}
    \label{fig:pred_rate}
\end{figure}

\paragraph{Multi-GPU instances.}
To demonstrate the generality of \Sys, we also evaluate \Sys on multi-GPU instances.
Unfortunately, we fail to collect meaningful multi-GPU spot instance traces on the cloud (e.g. {\tt p3.8xlarge} with 4 V100 GPUs) as they show extremely low availability recently. Instead, we propose to generate the 4-GPU instance based on the single GPU trace by accumulating every four preemption or allocation events. Each 4-GPU instance is allocated at the first allocation event and preempted at the last preemption event. In this way, multi-GPU instance trace will have higher GPU hours than the single GPU trace in total. For multi-GPU instances, we follow prior work~\cite{Megatron, DeepSpeed} using pipeline parallelism only for inter-nodes.
Figure~\ref{fig:multi_gpu} shows the training throughput and cost for different trace segments. Although our trace generation favors multi-GPU instances in theoretical availability, \Sys on single GPU instance still performs better in terms of both throughput and monetary cost. The major reason is that preempting one 4-GPU instance will interrupt 4 pipelines, significantly slowing down training. Besides, unutilized 4-GPU instances are also significant as it takes four times more GPUs to increase a new pipeline.

\subsection{Breakdown Analysis}
\paragraph{Look-ahead interval length.}
Figure~\ref{fig:ahead_ablation} shows the results of training GPT-2 with different numbers of look-ahead intervals on the $H_{A}D_{P}$ trace. Here \Sys looks back past 12 intervals and predicts the next $1, 4, 8, 12$, and $14$ intervals respectively. The results show that \Sys (Ideal) keeps improving by considering longer futures and achieves the best performance when looking ahead 12 intervals. It shows the benefits of \liveput-optimized configurations by considering future preemptions and allocations. 
On the other hand, \Sys exhibits a slightly different pattern, where its performance improves significantly by looking ahead 4 intervals compared with 1 interval ($1.8\times$). As \Sys looks ahead more intervals, the prediction error increases as we evaluated in \Cref{fig:arima_vs_others}. \Cref{fig:ahead_ablation} shows that \Sys can still yield significant improvement compared with looking ahead 1 interval, and achieves best performance by looking ahead 12 intervals. Overall, \Sys's throughput is $12.8\%$ lower than that of the ideal case. The result demonstrates that looking ahead longer can indeed help \Sys make more optimized decisions, and that there is still room to improve our availability predictor.

\paragraph{Prediction rate.} Figure~\ref{fig:pred_rate} shows the results of training GPT-2 with different prediction rates on the $H_{A}D_{P}$ trace. 
As the prediction rate decreases, so do the training throughput achieved by \Sys and \Sys (Ideal). Fortunately, the execution time of the liveput optimizer is much less than one minute, which allows \Sys to use a high prediction rate and optimize frequently (i.e., per minute) for better performance.

\begin{figure}[t]
    \centering
    \includegraphics[width=1\linewidth]{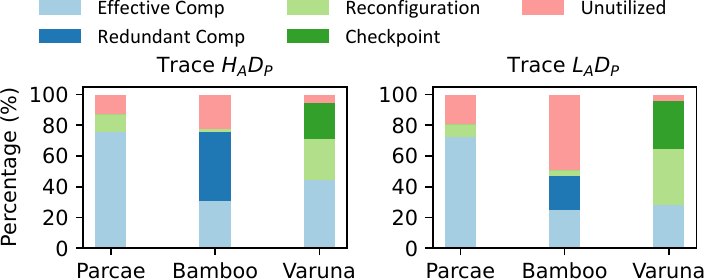}
    \vspace{\captionvspace}
    \caption{GPU hours breakdown of GPT-2 execution.}
    \label{fig:breakdown}
\end{figure}

\paragraph{GPU hours breakdown.}
To further understand the performance and drawbacks of different approaches, we breakdown the GPU hours of GPT-2 training into five components (Figure~\ref{fig:breakdown}). 
The results demonstrate that \Sys spends the majority of GPU hours performing effective computation (i.e., committed mini-batches).
In contrast, Bamboo spends more than $40\%$ GPU hours on redundant computation on $H_{A}D_{P}$, while wastes more than $50\%$ GPU hours on $L_{A}D_{P}$. 
Similarly, Varuna takes a long time to handle preemptions, including checkpointing and reconfiguration.
As a result, their unutilized parts are quite small compared with \Sys.
The results also align with the disadvantages we mentioned in \S\ref{eval:end_to_end}.

\paragraph{\Sys components analysis.}
Figure~\ref{fig:speedup_ablation} shows how each component contributes to the performance improvements, using GPT-2 as an example. We start from a checkpoint-based approach with throughput-optimized execution plans. 
By adding \SysPS and migration strategies, we improve the throughput by 13\%-67\%. Especially for trace $L_{A}D_{P}$ with low availability, it leaves little room for parallel configuration variation. When there are frequent preemption and allocation events, the migration allows training to make more progress than frequently triggering the costly reconfiguration. Finally, adopting \liveput optimized parallel configurations improves an additional $25.5\%$ over migration mechanisms.

\setcounter{figure}{12}
\begin{figure}[t]
    \centering
    \includegraphics[width=0.9\linewidth]{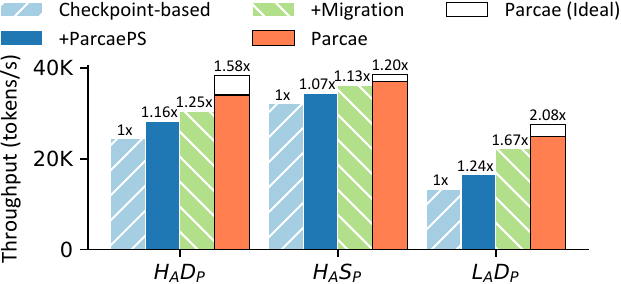}
    \vspace{\captionvspace}
    \caption{The decomposed throughput speedup on GPT-2.}
    \label{fig:speedup_ablation}
\end{figure}

\setcounter{figure}{13}
\begin{figure}[t]
    \centering
    \includegraphics[width=1\linewidth]{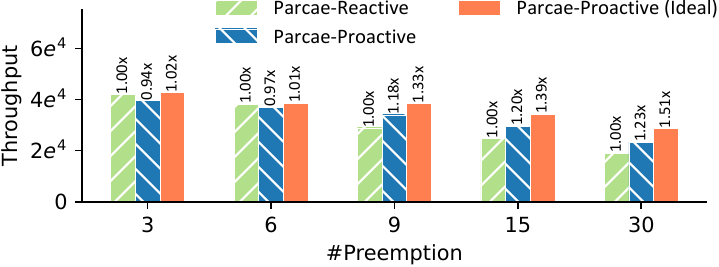}
    \vspace{\captionvspace}
    \caption{The throughput comparison between \Sys and \Sys-Reactive under different preemption intensity.}
    \label{fig:reactive}
\end{figure}

\setcounter{figure}{14}
\begin{figure*}[t]
    \centering
    \subfloat[Parallel configuration ($D\times P$) and average throughput inside each interval (i.e., 1 minute).]{
        \centering
        \includegraphics[width=0.81\linewidth]{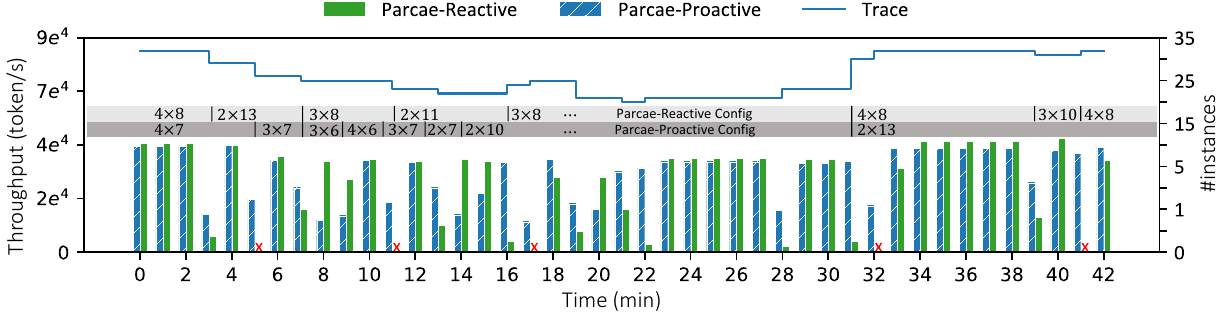}
        \label{fig:gpt_case}
    }
    \subfloat[Accumulated tokens.]{
        \centering
        \includegraphics[width=0.18\linewidth]{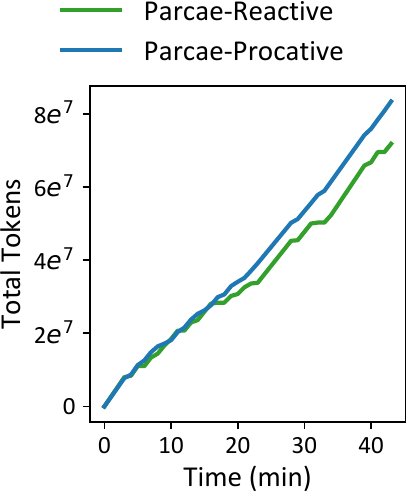}
        \label{fig:gpt_case_acc}
    }
    \vspace{\captionvspace}
    \caption{The comparison between \Sys-Reactive and \Sys-Proactive approaches for GPT-2 on $H_{A}D_{P}$ trace.}
    \label{fig:liveput_tpt}
\end{figure*}

\setcounter{figure}{15}
\begin{figure}[t]
    \centering
    \includegraphics[scale=0.5]{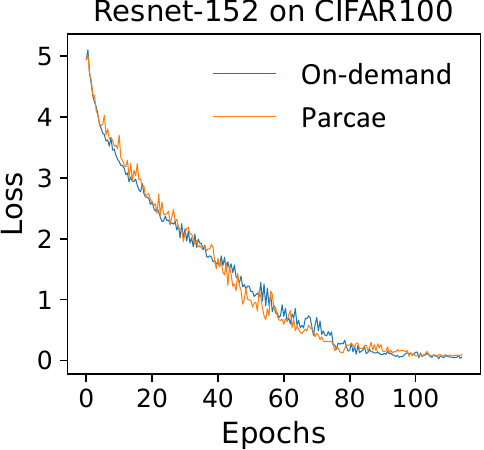}
    \vspace{\captionvspace}
    \caption{The loss curve of ResNet-152 on CIFAR100.}
    \label{fig:convergence}
\end{figure}

\subsection{Proactive v.s. Reactive}
\paragraph{Preemption Tolerance.} 
We evaluate the performance of \Sys (i.e., \Sys-Proactive) and \Sys-Reactive with GPT-2 on a synthetic preemption trace~\Cref{fig:reactive}.
The auxiliary baseline (i.e., \Sys-Reactive) is created by disabling the liveput optimization in \Sys and only enabling the parallelization adaptation mechanism (\S\ref{sec:fault}). \Sys-Reactive can be classified as a throughput-optimized system and used to highlight the advantages of our proactive, liveput-based approach. We generate the synthetic trace from the $H_AS_P$ trace by scaling the number of preemption events from 3 to 30 within one hour. The performance gap between \Sys-Reactive and \Sys-Proactive becomes larger as the preemption intensity increases, showing that our proactive approach can be more effective for scenarios with more frequent preemptions.

\paragraph{Case study.}
As a case study, we compare the \liveput-optimized \Sys with \throughput-optimized \Sys-Reactive in detail using GPT-2 and partial $H_{A}D_{P}$ trace.
Figure~\ref{fig:gpt_case} shows each interval's instance availability, parallel configuration ($D\times P$), and average throughput as time elapses.
We observe that for intervals with stable availability, \Sys-Reactive can select configurations with relatively higher throughput.
However, greedily selecting execution plans that maximize throughputs suffers when preemptions or allocations happen because it neglects high reconfiguration costs. It can barely make training progress when the available instances frequently change. In contrast, \Sys carefully chooses parallel configurations by considering the future instance availability and adapting efficient migration strategies accordingly to ensure high training efficiency while mitigating expensive reconfiguration. 
For example, in the first 8 intervals, \Sys selects a pipeline depth of 7 and avoids changing pipeline depth as \Sys-Reactive does (e.g., 8 and 13). Although resulting in some unused instances, the progress made is still larger than running with \Sys-Reactive because of its reconfiguration overheads.
Similar observation exists in the last 10 consecutive intervals, where \Sys maintains the same parallel configurations but leverages lightweight inter- and intra-stage migrations to adapt to dynamic preemptions and allocations.
As a result, \Sys achieves 16\% more accumulated tokens within 40 minutes (Figure~\ref{fig:gpt_case_acc}).

\subsection{Convergence Preservation}
Figure~\ref{fig:convergence} visualizes the convergence curve of training loss between \Sys running on spot instances and the baseline on on-demand instances. We observe that both convergence rates are very close, and \Sys reaches the same training loss of 0.058 as the baseline after training for 110 epochs. This verifies that the \Sys design and implementation align with the model convergence.
\section{Related Work}
\label{sec:related}

\paragraph{Dynamic preemptible instances.}
There is a trend of using preemptible instances on modern clouds for cheap service. 
Tributary~\cite{harlap2018tributary} studies the latency issues from preemptions and proposes to switching preemptible offerings from clouds with different preemption likelihoods. 
BurScale~\cite{baarzi2019burscale} employs autoscaling to handle transient queuing in web service traffic. 
SciSpot~\cite{kadupitiya2022scispot} presents a reliability model for temporally constrained preemptions to optimize the job scheduling for scientific computing. 
HotSpot~\cite{hotspot} transparently migrates spot VMs in lower price and achieves higher cost-efficiency.
Snape~\cite{yang2023snape} improves spot resources' availability by dynamically mixing on-demand VMs with spot eviction predictions. 
SpotServe~\cite{asplos24spotserve} realizes fast and reliable serving of LLMs on cheap preemptible instances with dynamic reparallelization and optimal context migration.
Prior work has demonstrated the cost benefits of spot instances in cloud computing and motivates the following related research.

\paragraph{Preemptible distributed DNN training.}
Recently, using preempible instances for machine learning
tasks is becoming popular as they are much more cost effective, like what is done in Varuna~\cite{athlur2022varuna} and Bamboo~\cite{bamboo}. 
SageMaker in AWS~\cite{sagemaker} automatically pauses the training job when a spot instance is interrupted and resumes from the checkpoint in S3 if the spot instance becomes available again.
CM-DARE~\cite{li2020characterizing} analyzes distributed training under transient cloud GPU servers and provides a performance modeling methodology. SpotTune~\cite{li2020spottune} leverages spot instances to paralleize hyper-parameter tuning for ML models. 
SkyPilot~\cite{286502} migrates training workload to spot resources from other clouds and relaunch jobs using the periodical checkpoint from cloud storage.
These approaches make meaningful explorations in this direction but are still suffering from the limited performance due to preemptions.

Oobleck~\cite{oobleck2023sosp} and Gemini~\cite{gemini2023sosp} are concurrent works for quick failure recovery in distributed DNN training. Oobleck introduces pipeline reinstantiation with pre-computed pipeline templates. Gemini uses in-memory checkpoints and orchestras checkpoint traffic schedule. Both are reactive approaches. Besides, Gemini targets dedicated instances and relies on high network bandwidth to reduce checkpointing time, while the bandwidth is low for spot instances.

\section{Conclusion}
In this work, we present an efficient distributed training system over spot instances, \Sys. The key idea is to proactively adjust the parallelization strategy using a novel metric, liveput, considering both training throughput and instance availability. With holistic system mechanisms and implementation optimizations, \Sys significantly outperforms checkpoint- and redundancy-based solutions in evaluations.

\section*{Acknowledgement}
We thank the anonymous reviewers and our shepherd, Le Xu, for their comments and helpful feedback.
This material is based upon work supported by NSF awards CNS-2147909, CNS-2211882, and CNS-2239351, and awards from Amazon, Cisco, Google, Meta, Oracle, Qualcomm, and Samsung.

\bibliographystyle{plain}
\bibliography{myref}


\clearpage
\appendix 

\section{Addition Details of Migration Costs}
\label{appendix:mig_cost}

\Cref{table:time} lists detailed costs of migrations and their magnitudes. All of them are profiled multiple times and averaged over five DNN models (see \Cref{eval:models}).

\section{Additional Details of ARIMA}
\label{appendix:arima}

The ARIMA time-series forecasting algorithm is sensitive to trivial perturbations in inputs, which may impede its understanding of essential patterns from previous instance history. We introduce a few optimizations to ensure its predictions are faithful. First, we flatten random spikes that last for only 1-2 intervals in previous instance history, since such trivial noise will likely cause abrupt rise and falls in prediction. ARIMA also likes to simulate the tendency of the entire input curves. When input curves have multiple "hops", we ensure that ARIMA only learns from the most recent variations that are indeed beneficial for prediction. Second, though ARIMA can accurately capture intermediate fluctuations, its prediction can be so steep that it easily hits the upper and lower boundaries of available instances on intervals of sudden increase and decrease. To do so, we set upper and lower boundaries to limit the predicted curves based on observations of all spot instance traces we have. Additionally, our empirical study on traces indicate most intervals have a limitation on the extent of growth. Thus, we would also apply such constraints on predictions. We also apply additional penalty to flatten excessively steep predictions such as their predictions follow the essential patterns of AWS traces. We take care to reset ARIMA mispredictions when the generation deviates seriously from the input. With these rules and modifications, we ensure the ARIMA model can sufficiently describe future scenarios by learning from the past history.        

\section{Additional Experimental Details}
\label{appendix:exp_detail}

\begin{table}[t]
\centering
\caption{Overview of the five DNNs evaluated.}
\footnotesize
\label{eval:models}
\begin{tabular}{ l|lll }
\hline
Model & {mini-batch} & {micro-batch} & Dataset \\
\hline
ResNet-152~\cite{resnet}   & 2048  & 32     & CIFAR-100~\cite{cifar100}  \\
VGG-19~\cite{VGG}         & 2048  & 32      & CIFAR-100~\cite{cifar100}  \\
\cline{1-4}
BERT-Large~\cite{bert}    & 1024 & 8 & WikiText-2~\cite{wikitext2} \\
GPT-2 (1.5B)~\cite{gpt2} & 128  & 1 & WikiText-2~\cite{wikitext2} \\
GPT-3 (6.7B)~\cite{gpt3}  & 64 & 1 & WikiText-2~\cite{wikitext2} \\
\hline
\end{tabular}
\end{table}

\begin{table}[t]
\centering
\footnotesize
\caption{Migration costs in our experiments on AWS.}
\vspace{-1em}

\label{table:time}
\begin{tabular}{ l|cc}
 \hline
 Cost Terms & Magnitude ($s$) & Interfering Factor  \\
 \hline
 Start process & $< 1$ & \multirow{3}{*}{Instance state} \\
 Rendezvous & $0 \sim 10$ & \\
 Init CUDA context  & $0\sim 10$ & \\
 \hline
 Load data & $0\sim 10$ & Dataset \\
 \hline
 Build model & $0\sim 10$ & \multirow{2}{*}{Model, Configuration} \\
 Update comm. groups & $0\sim 20$ & \\
 \hline
 \multirow{2}{*}{Model states transfer} & \multirow{2}{*}{$0\sim 60$} & Model, Configuration \\
 & & Preemption Scenario \\
 \hline
\end{tabular}
\end{table}

\subsection{End-to-End Evaluation Setting}
We select five popular DNNs for various applications and summarize them in Table~\ref{eval:models}. For all the models, we used Adam optimizer with half precision (i.e., FP16) for training. 

\paragraph{Parallel Configuration.} \Sys and Varuna will adjust parallel configurations according to instance availability during training. The parallel configuration of \Sys is decided by migration manager, while it is decided by job morphing for Varuna. We follow the settings of Vauna and first run a one-time profiling to collect primitive parameters of the hardward and the DNN model. Varuna will automatically decide the optimal parallel configuration considering DNN models and number of availability instances. Table~\ref{eval:bamboo} summarizes the parallel configurations used for Bamboo in our evaluation. Bamboo maintains a fixed pipeline depth and its redundant computation consumes a huge amount of memory. For different models, we tuned the number of pipeline stages and partitions to find an optimal parallel configuration for Bamboo. We find it requires at least 20 stages for Bamboo to run GPT-3 even with activation checkpointing~\cite{tq_sublinear} enabled, and Bamboo performs best for $P = 23$. 

\begin{figure}
    \centering
    \includegraphics[width=\linewidth]{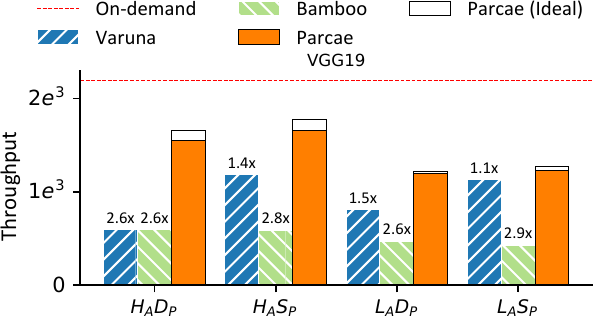}
    \caption{Training throughput comparison of VGG19 among existing frameworks and Parcae on four traces. The dotted on-demand line shows the best throughput with on-demand instances. The numbers over the bars represent the speedup of Parcae over Varuna and Bamboo respectively.}
    \label{fig:vgg19}
\end{figure}

\paragraph{VGG Results}
\Cref{fig:vgg19} shows the end-to-end evaluation results of VGG19. \Sys significantly outperforms Varuna and Bamboo, except for trace $L_{A}S_{P}$, where Varuna achieves comparable performance with \Sys. We move these results in the appendix due to the limited page space.

\begin{table}[h]
\centering
\caption{The parallel configuration of Bamboo in evaluation.}
\label{eval:bamboo}
\begin{tabular}{ l|cc }
\hline
Model & D & P \\
\hline
ResNet-152~\cite{resnet}   & 8 & 4 \\
VGG-19~\cite{VGG}         & 8 & 4 \\
\hline
BERT-Large~\cite{bert}    & 4 & 8 \\
GPT-2 (1.5B)~\cite{gpt2} & 2 & 16 \\
GPT-3 (6.7B)~\cite{gpt3}  & 1 & 23 \\
\hline
\end{tabular}
\end{table}

\subsection{\Sys Components Evaluation}

\begin{figure}[t]
    \centering
    \subfloat[The accuracy of cost estimator.]{
        \centering
        \includegraphics[width=0.45\linewidth]{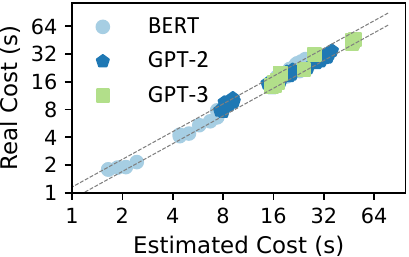}
        \label{fig:cost_est}
    }
    \subfloat[The cost of migration advisor.]{
        \centering
        \includegraphics[width=0.5\linewidth]{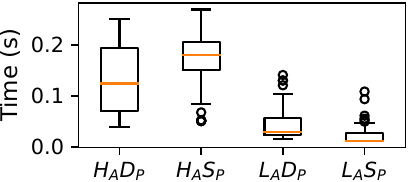}
        \label{fig:opt_time}
    } 
    \caption{(a) Comparison between the estimated and actual reconfiguration time for different models. (b) Optimization time of looking ahead 12 intervals for GPT-2.}
    \label{fig:component}
\end{figure}

\paragraph{Cost Estimation Accuracy.}
The cost estimator estimates migration cost for different preemption scenarios and parallel configurations. An accurate estimator is important for accurate \liveput optimization. We compare the estimated migration cost predicted by cost estimator with the real migration time measured by actual executions. Figure~\ref{fig:cost_est} shows the results for different DNN models. The dashed lines indicate a relative difference of $-15\%$ and $15\%$ between real and estimated migration cost, respectively. The results demonstrate that our cost estimator is appropriate to evaluate the migration cost for different preemption scenarios and models. 

\paragraph{Optimization Cost.}
\SysSched periodically runs online \liveput optimization to suggest the parallel configuration for the next interval. We evaluate the optimization time it takes to look ahead 12 intervals for one run on one CPU machine. Figure~\ref{fig:opt_time} shows the results of GPT-2 on different trace segments. Overall, one optimization takes less than 0.3 seconds, which is negligible compared with interval length. Therefore, the \liveput optimization will not delay the training process.

\end{document}